\pdfoutput=1
\documentclass[numberedappendix]{emulateapj}

\usepackage{amsmath,amssymb}
\usepackage{float}
\usepackage{CJK}
\usepackage{color}

\def\gtrsim{\mathrel{\hbox{\rlap{\hbox{\lower4pt\hbox{$\sim$}}}\hbox{\raise2pt\hbox{$>$}}}}}

\newcommand{\kms}{km~s\ensuremath{^{-1}}}

\newcommand{\mbulge}{\ensuremath{M_\mathrm{bulge}}}

\newcommand{\mbh}{\ensuremath{M_\mathrm{\rm{BH}}}}
\newcommand{\mdm}{\ensuremath{M_\mathrm{\rm{DM}}}}

\newcommand{\mgal}{\ensuremath{M_{\mathrm{\rm{BH}}}-M_{\mathrm{bulge}}}}
\newcommand{\msigma}{\ensuremath{M_{\mathrm{\rm{BH}}}-\sigmastar}} 
\newcommand{\mvc}{\ensuremath{M_{\mathrm{BH}}-V_\mathrm{c}}} 
 
\newcommand{\msun}{\ensuremath{M_{\odot}}}

\newcommand{\sigmastar}{\ensuremath{\sigma_{\ast}}}

\newcommand{\Vc}{V_\mathrm{c}}
\newcommand{\degree}{^\circ}
\newcommand{\HI}{H~{\small I}~}

\def\lax{{$\mathrel{\hbox{\rlap{\hbox{\lower4pt\hbox{$\sim$}}}\hbox{$<$}}}$}}
\def\gax{{$\mathrel{\hbox{\rlap{\hbox{\lower4pt\hbox{$\sim$}}}\hbox{$>$}}}$}}

\shorttitle{{\it $\mvc$} RELATION}
\shortauthors{Sun et al.}

\begin{document}
\title{Refining the $\mvc$ scaling relation with HI rotation curves of water megamaser galaxies}
\author{Ai-Lei Sun\altaffilmark{1}, Jenny E. Greene\altaffilmark{1,5}, 
C. M. Violette Impellizzeri\altaffilmark{2,6}, Cheng-Yu Kuo\altaffilmark{3}, 
James A. Braatz\altaffilmark{2}, Sarah Tuttle\altaffilmark{4}}

\altaffiltext{1}{Department of Astrophysics, Princeton University, Princeton, NJ 08540, USA}
\altaffiltext{2}{National Radio Astronomy Observatory, 520 Edgemont Road, 
Charlottesville, VA 22903, USA}
\altaffiltext{3}{Academia Sinica Institute of Astronomy and Astrophysics, P.O. Box 23-141, Taipei 10617, Taiwan}
\altaffiltext{4}{McDonald Observatory,The University of Texas, Austin, Texas 78712, USA}
\altaffiltext{5}{Alfred P. Sloan Fellow}
\altaffiltext{6}{Joint Alma Office, Alsonso de Cordova 3107, Vitacura,  Santiago, Chile}

\begin{abstract}

Black hole - galaxy scaling relations provide information about
the coevolution of supermassive black holes and their host
galaxies. We compare the black hole mass - 
circular velocity ($\mvc$) relation with the 
black hole mass - bulge stellar velocity dispersion ($\msigma$) relation, to see whether the scaling
relations can passively emerge from a large number of mergers, or
require a physical mechanism, such as feedback from an active nucleus.
We present VLA \HI observations of five galaxies, including three
water megamaser galaxies, to measure the circular velocity.  Using twenty-two
galaxies with dynamical $\mbh$ measurements and $\Vc$ measurements extending to
large radius, our best-fit $\mvc$ relation, $\log \mbh = \alpha +
\beta \log (\Vc/ 200 ~\rm{km~ s^{-1}})$, yields $\alpha =
7.43^{+0.13}_{-0.13}$, $\beta = 3.68^{+1.23}_{-1.20}$, and intrinsic
scatter $\epsilon_{int}=0.51^{+0.11}_{-0.09}$. 
The intrinsic scatter may well be higher than 0.51, as we
take great care to ascribe 
conservatively large observational errors. We find comparable scatter in the $\msigma$
relations, $\epsilon_{int} =0.48^{+0.10}_{-0.08}$, while pure merging
scenarios would likely result in a tighter scaling with the dark halo
(as traced by $\Vc$) than baryonic ($\sigma_*$) properties. Instead,
feedback from the active nucleus may act on bulge scales to tighten
the $\msigma$ relation with respect to the $\mvc$ relation, as
observed.
\end{abstract}

\section{Introduction}

The observed scaling relations between supermassive black hole (BH)
mass and properties of the host galaxy, intensively studied over the
past decade, suggest that black hole growth is tied to the growth of the
surrounding host galaxy. These galaxy properties include the
bulge/spheroid stellar velocity dispersion $\sigmastar$
\citep[e.g.,][]{Ferrarese00, Tremaine02, Gueltekin09, Beifiori12,
  McConnell12}, the mass and luminosity of galaxy
bulges \citep[e.g.,][]{Marconi03,Haering04, McConnell12}, and the circular velocity $\Vc$ \citep[e.g.,][]{Ferrarese02, Kormendy11, Beifiori12}, which is the rotation velocity measured at large radius to probe the dark matter halos potential. 
 It is
intriguing that these power-law relations, especially the $\msigma$
relation, hold over several orders of magnitude in BH mass with small
scatter, even though the black hole accounts for only a few
thousandths of the mass of the galaxy \citep[e.g.,][]{Haering04}.

There are a wide array of theories attempting to explain the BH-galaxy
scaling relations \citep[e.g.,][]{Silk98, CiottiOstriker01,
  Murrayetal05, Hopkinsetal06, Peng07}. Two of the most popular
models include variants of feedback from active galactic nuclei
(AGN) and scenarios in which merging alone can lead to BH-galaxy
scaling laws. In AGN feedback models, the central BH accretes mass and
grows until it is massive enough to expel gas
from the galaxy potential well and quench its own growth. 
BH growth in this picture is regulated by the depth of the galaxy potential well
\citep{Silk98, Fabian99, DiMatteo05, Robertson06a}. On the other
hand, the pure merging scenario suggests that the correlation between
linear quantities, for example the BH mass $\mbh$ and the halo mass
$\mdm$, can emerge from a large number of mergers based on the central
limit theorem, even without a physical mechanism linking the two
\citep{Peng07,Hirschmann10, Jahnke11}. 

Although both the feedback and merging phenomena may occur in galaxy
evolution, it is unclear whether either of the mechanisms is essential
in establishing the scaling relations. The most important physical
scale for feedback is also a matter of debate \citep{BoothandSchaye10,
  Debuhr10}. Furthermore, we do not know how the AGN
  output couples to the gas, whether via thermal energy \citep{Silk98}
  or momentum \citep{Ostriker10}, nor do we know the average
efficiency of the feedback.

Therefore, empirical evidence that distinguishes the relative
importance of different physical processes in establishing the scaling
relations is key to constructing the coevolution history of black
holes and galaxies.  In this paper, we investigate the origin of
BH-galaxy scaling relations by comparing the $\mvc$ relation to the
$\msigma$ relation.  Circular velocity $\Vc$ is a good indicator of
dark matter halo mass, and velocity dispersion $\sigmastar$ serves as
its counterpart on bulge scales. While some AGN feedback scenarios
\citep[e.g.,][]{Debuhr10} suggest that BH mass will be most tightly
linked to baryons (as opposed to dark matter), a pure merging scenario suggests that the
$\mbh-\mdm$ (or $\mvc$) relation should be the cleanest and tightest
relation, as it is free from the baryonic physics (e.g., star
formation) that occurs during merging.  Comparison of the two
relations, especially their scatter, can help determine the mechanism
that drives BH-galaxy coevolution \citep{Ferrarese02, Novak06,
  Kormendy11}.

\citet{Ferrarese02} first proposed that BH mass may correlate with
dark matter halo mass, based on the $\msigma$ relation and the
correlation between $\sigmastar$ and $\Vc$. Later, a number of papers
\citep{Pizzella05,Courteau07,Ho07} pointed out that the $\sigmastar-\Vc$
relation depends on surface brightness, light concentration, and
morphology, and suggested that the $\msigma$ relation, not the $\mvc$ relation, is
most fundamental. \citet{Kormendy11} compiled a sample of 25 galaxies
with both dynamical BH mass measurements and $\Vc$ from spatially
resolved rotation curves. From this direct $\mvc$ correlation they
concluded that the dark matter halo mass alone cannot determine the BH
mass, given that the BH mass can range from $<10^3-10^6$~\msun\ at a
circular velocity of $120 ~\rm{km~s^{-1}}$ \citep[for a different
view see][]{Volonteri11}.  \citet{Beifiori12} also found the
  scatter in the $\mvc$ relation to be about twice as large as that in the
  $\msigma$ relation using a large sample of $\mbh$ upper limits from {\it
    Hubble Space Telescope} spectra \citep{Beifiori09} and $\Vc$ from unresolved H
  {\small I} line-width measurements. For a comprehensive review,
see \citet{Kormendy13}.
 
In this work, we aim to refine the $\mvc$ statistics.  We start with
the most up-to-date galaxy sample with dynamically measured black hole
masses, and study all of these that also have spatially resolved
circular velocity measurements, predominantly disk
  galaxies observed with H~{\small I}.  We present five new \HI
rotation curves measured with the Karl G. Jansky Very Large
Array\footnotemark[1] (VLA) during the Expanded Very Large Array 
(EVLA) construction period. Three of these galaxies have
dynamical $\mbh$ measured from the kinematics of water megamasers in
sub-pc disks \citep{MCPI,Greene10,Kuo11}.  With BH mass errors smaller
than 11 percent, these megamaser measurements are especially useful in
constraining the intrinsic scatter of the $\mvc$ relations. In total
our sample contains thirty-three galaxies. We assign $\Vc$
to all galaxies in a consistent way and investigate $\Vc$ reliability as a
function of the spatial extension of the rotation curve.  Using only
reliable $\Vc$ measurements at large radius, we quantify the $\mvc$
correlation and compare it with the $\msigma$ relation. We investigate
whether \mbh\ is correlated more tightly with $\Vc$ or $\sigmastar$,
to discriminate AGN feedback scenarios from those in which merging
alone establishes the scaling relations.  \footnotetext[1]{Operated by
  The National Radio Astronomy Observatory, a facility of the National
  Science Foundation operated under cooperative agreement by
  Associated Universities, Inc.}

\section{\HI Observations}
\label{sec:observation}

We observed 5 spiral galaxies in \HI with the VLA.  The observations were
taken in the L-band and the C configuration under project 10B$-$220
between October and December 2010 (for details see Table \ref{tab:obsquality}).  The
observations used dual circular polarizations and a single spectral
window with 256 channels across the 4MHz (852 km~s$^{-1}$) bandwidth,
giving a channel width of $3.3~\rm{km~s^{-1}}$. Twenty-one edge
channels on each side were flagged due to a higher noise
level. Removing these channels did not affect our results since none of
the sources had emission in these parts of the band. The full width at
half power of the primary beam is $32\arcmin$, and the synthesized
beam size ranges from $23\arcsec$ to $52\arcsec$ depending on the
source declination and number of antennas used (Table \ref{tab:image quality}).

The duration of each observation was 5 hours and the on-source time
was $\sim 200$ minutes per source. During a track, the gain calibrator
was observed for $3-6$ minutes every $20-27$ minutes, while the
flux calibrator was observed once, for $15-23$ minutes. The
observations were carried out during the EVLA upgrade phase and
some of the L-band receivers were not yet installed. 
For the observations of NGC 1194, NGC 2748, and UGC 3789, four antennas 
were not in the array, three were missing during the NGC 7582 observations, 
and two during the NGC 2960 track. 
One additional antenna was flagged out of the data set for NGC 1194, NGC 2748, NGC
7582, and UGC 3789 for high noise levels.  
The total number of antennas used was between 22
and 25, as listed in Table \ref{tab:obsquality}. Radio Frequency
Interference (RFI) was found in three observations, and the
contaminated data were excluded, which accounts for only a few percent
  of the data.

Data calibration and image processing were carried out with the
Common Astronomical Software Applications (CASA) package. Time
averaging of 10 seconds and on-line flagging were applied before
calibration. The amplitude loss due to time averaging is less
than 1$\%$ according to the VLA Observational Status Summary. The
antenna position correction, antenna-based delays, bandpass,
phase gain, amplitude gain and flux scale calibrations were carried
out consecutively.  The minimum acceptable signal-to-noise ratio (SNR)
for the bandpass and gain calibration tables were all set to 3. The
bandpass solutions were solved for each channel, which gave stable
solutions with amplitude variations  $\lesssim 1 \%$. The phase gain and
amplitude gain were solved for each calibrator scan.  Gain elevation
curves were applied to correct for elevation-dependent amplitude gain
due to antenna deformation.  Data weighting was not used throughout
the calibration process, since the VLA does not monitor system
temperature. The two polarizations were calibrated separately and
combined in the imaging process.

For the \ion{H}{1} emission line analysis, the continuum was estimated
from line-free channels with more than 20 channels on each side and
subtracted in the $uv$ plane with the task $\it{uvcontsub}$.  A
constant continuum level was used for all galaxies but NGC 1194 and
NGC 2960.  In these two galaxies, an adapted linear spectrum was used
in order to subtract a nearby bright source with an inclined
spectrum. The nearby bright continuum sources were well-subtracted at
this stage, and the contamination from their side-lobes was negligible.

For imaging we used the CASA task $\it{clean}$ in $\it{velocity}$
mode. This mode corrects for the doppler shift due to rotation of the
Earth. In this process the channels were also regridded linearly into
wider spectral bins - the image planes. The width of the image planes
were chosen to be 10 km~s$^{-1}$ for higher signal-to-noise (SNR)
observations (NGC 2748 and NGC 7582), and 20 km~s$^{-1}$ for lower SNR
observations (NGC 1194, NGC 2960, and UGC 3789). The
  SNR ratios for each data cube, defined as the peak intensity per channel divided by the RMS
  noise, are listed in Table \ref{tab:image quality}. For NGC 2748 and
NGC 7582 we applied Briggs weighting with the parameter {\it robust}
set to 0.5, and natural weighting was applied for NGC 1194, NGC 2960,
and UGC 3789. All images were cleaned to the 3 $\sigma$ level.  All
velocities in this paper use the optical convention and the
 Local Standard of Rest Kinematic (LSRK) reference
  frame. Position angles (P.A.) in this paper are measured east of
  north.
  
\section{Analysis}
  In this section we describe the procedures that we used to extract velocity
  fields from the \HI data cubes and measure rotation curves and
  $\Vc$. The \ion{H}{1} properties of individual galaxies
  are discussed in Appendix \ref{sec:HI_prop}.

\subsection{Non-parametric moment-0 and moment-1 maps}
\label{sec:moment maps}

\begin{figure*}[h]
	\centerline{
	\vbox{
	\hbox{
	\hskip -0.15in
	\includegraphics[bb = 25 400 500 750,clip, scale=0.5]{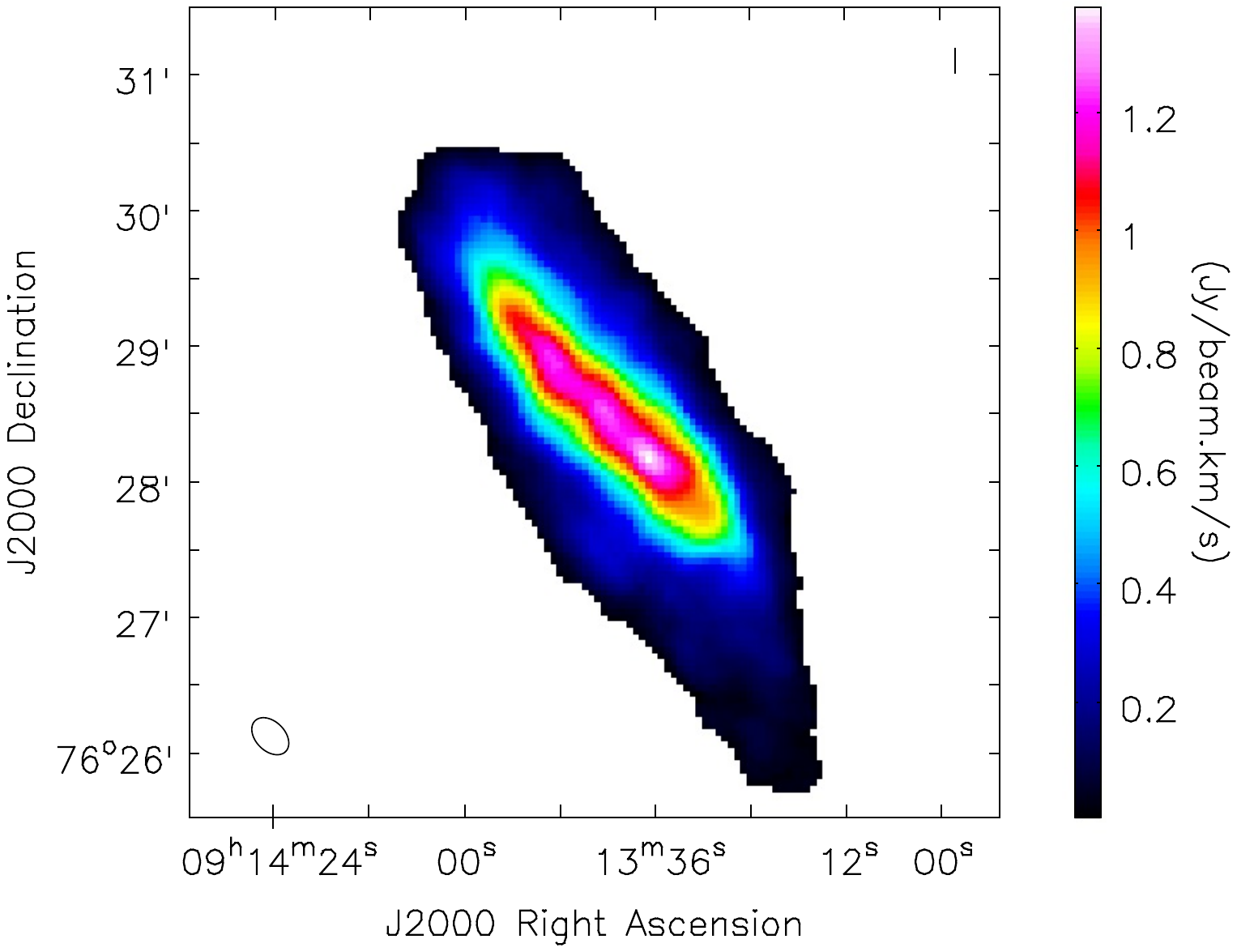}
	\hskip -2mm
	\includegraphics[bb = 25 400 500 750,clip, scale=0.5]{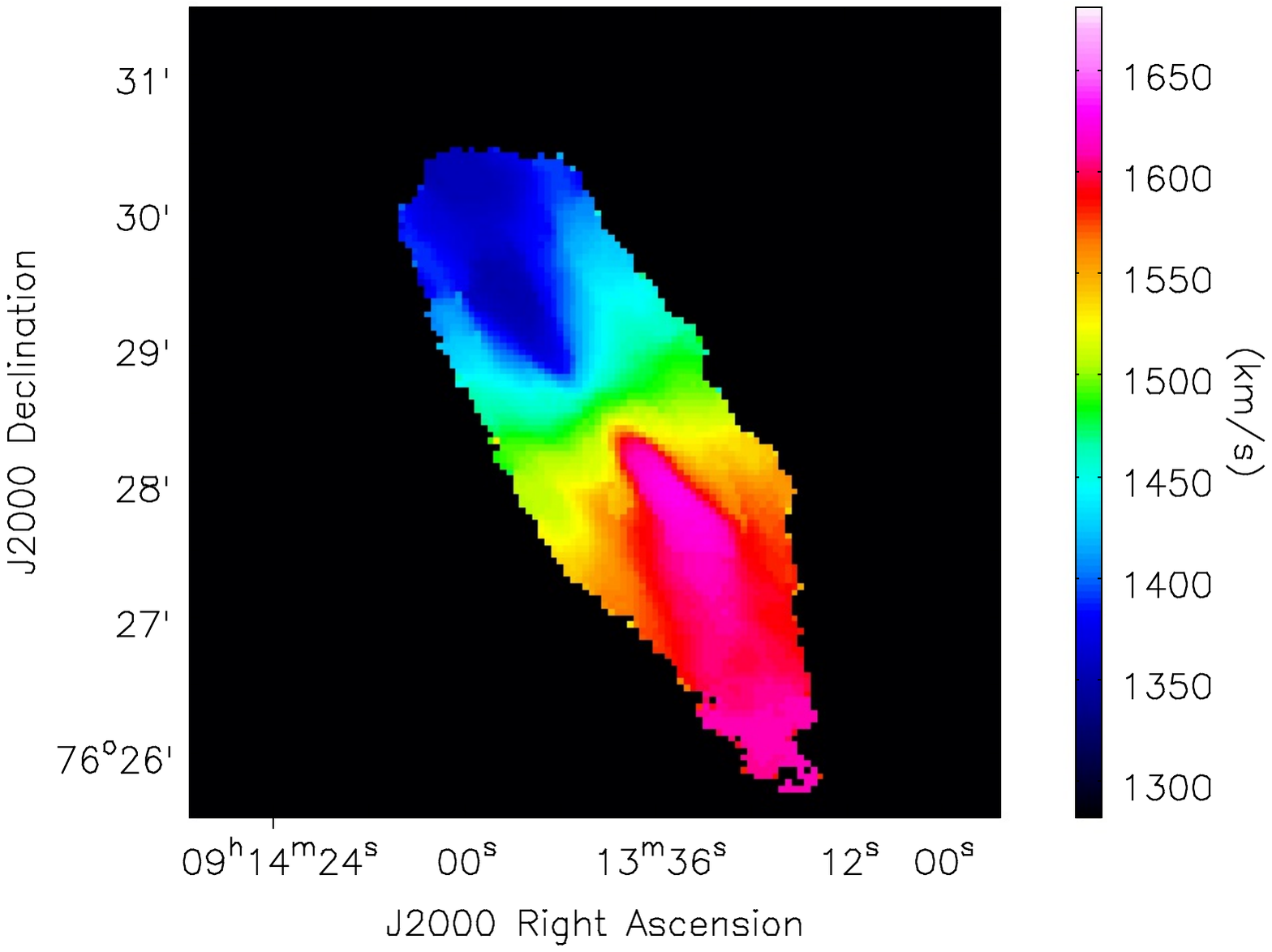}
	}
	}}
	\caption{
          Left: The NGC 2748 moment-0 map made by the CASA {\it
            image.moments} tool. The masking is described in \S
          \ref{sec:moment maps}. We mask out the pixels below 3 $\sigma$
          in a map that is smoothed spatially and over two velocity
          channels ($20 ~\rm{km~s^{-1}}$).  We then construct the moment-0 map from
          the original data using this mask. The ellipse in the bottom left corner represents the beam. Right: NGC 2748 Gaussian/Gauss-Hermite fitted
          velocity field. Velocity is in optical LSRK.}
        \label{fig:ngc2748}
\end{figure*}

\begin{figure*}[h]
	\centerline{
	\vbox{
	\hbox{
	\hskip -0.15in
	\includegraphics[bb = 25 400 500 750,clip, scale=0.5]{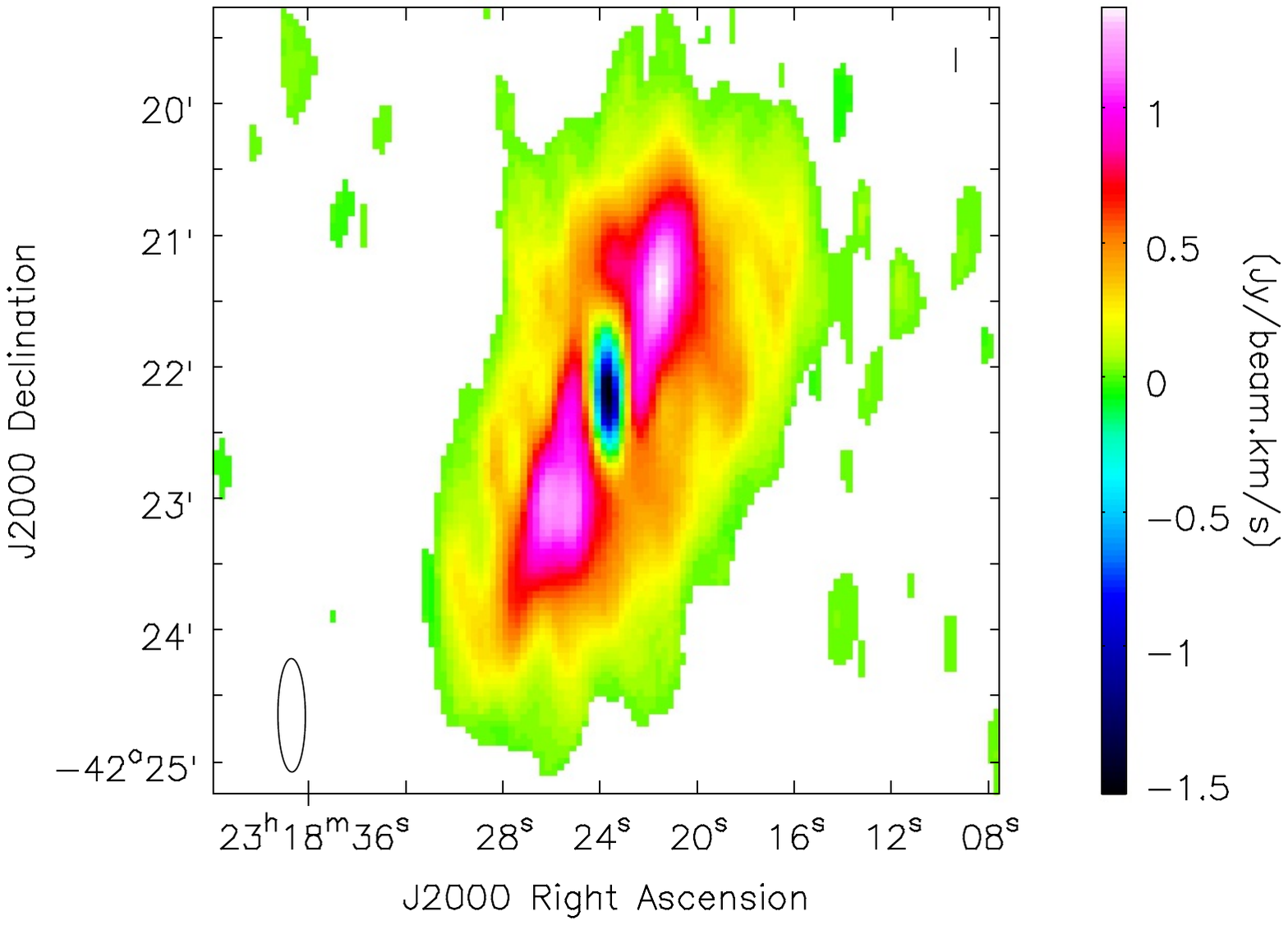}
	\hskip -1mm
	\includegraphics[bb = 25 400 500 750,clip, scale=0.5]{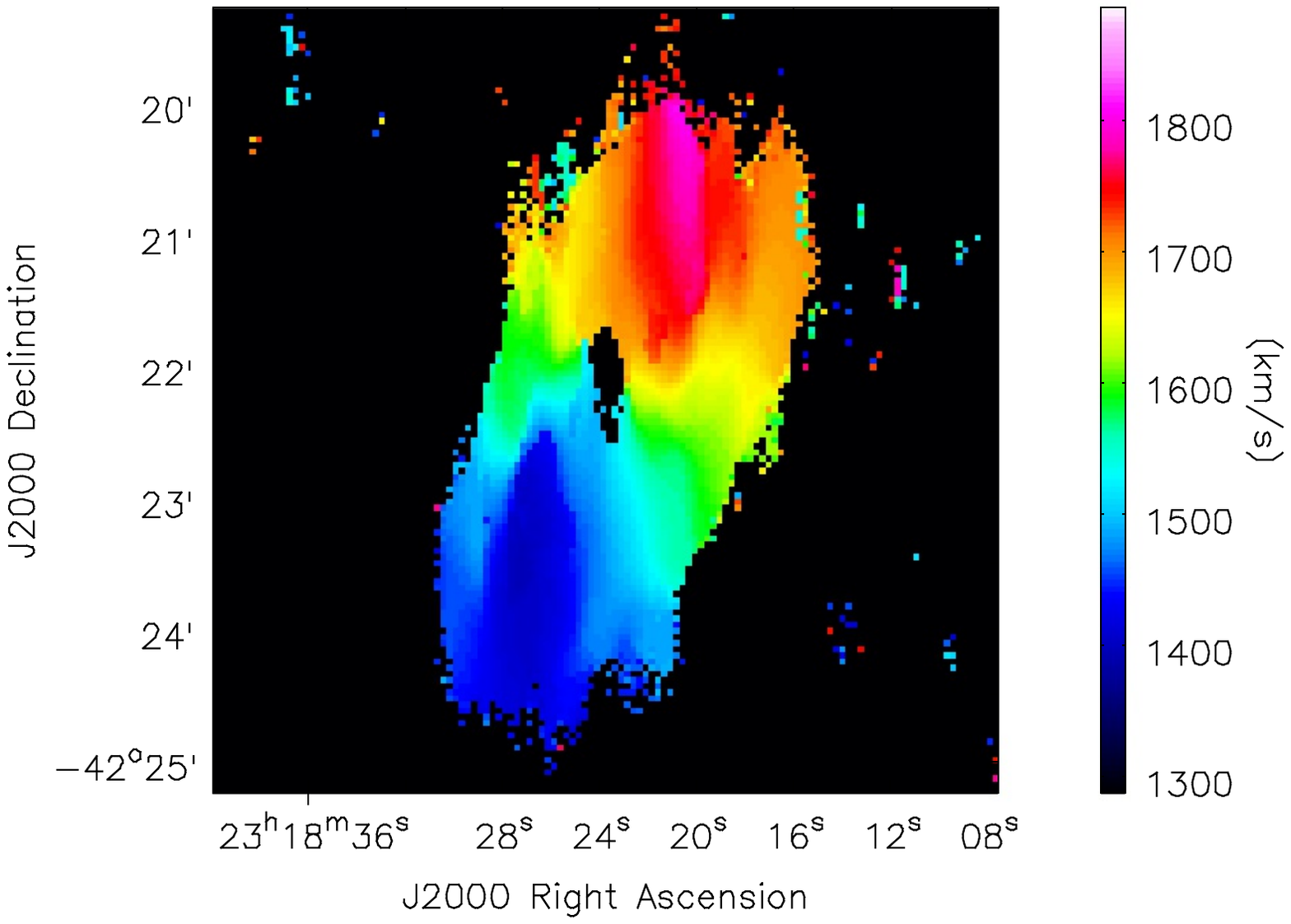}
	}
	}}
	\caption{
          Left: The NGC 7582 moment-0 map. The masking is described in
          \S \ref{sec:moment maps}. We mask out the pixels below 3 $\sigma$
          in a map that is smoothed spatially and over two velocity
          channels ($20 ~\rm{km~s^{-1}}$).  We then construct the moment-0 map from
          the original data using this mask. The ellipse in the bottom left corner represents the beam.  Right: The NGC 7582
          Gaussian/Gauss-Hermite fitted velocity field. The central
          region is masked because of the absorption feature.}
        \label{fig:ngc7582}
\end{figure*}

\begin{figure*}[h]
	\centerline{
	\vbox{
	\hbox{
	\hskip -0.2in
	\includegraphics[bb = 25 400 500 750,clip, scale=0.5]{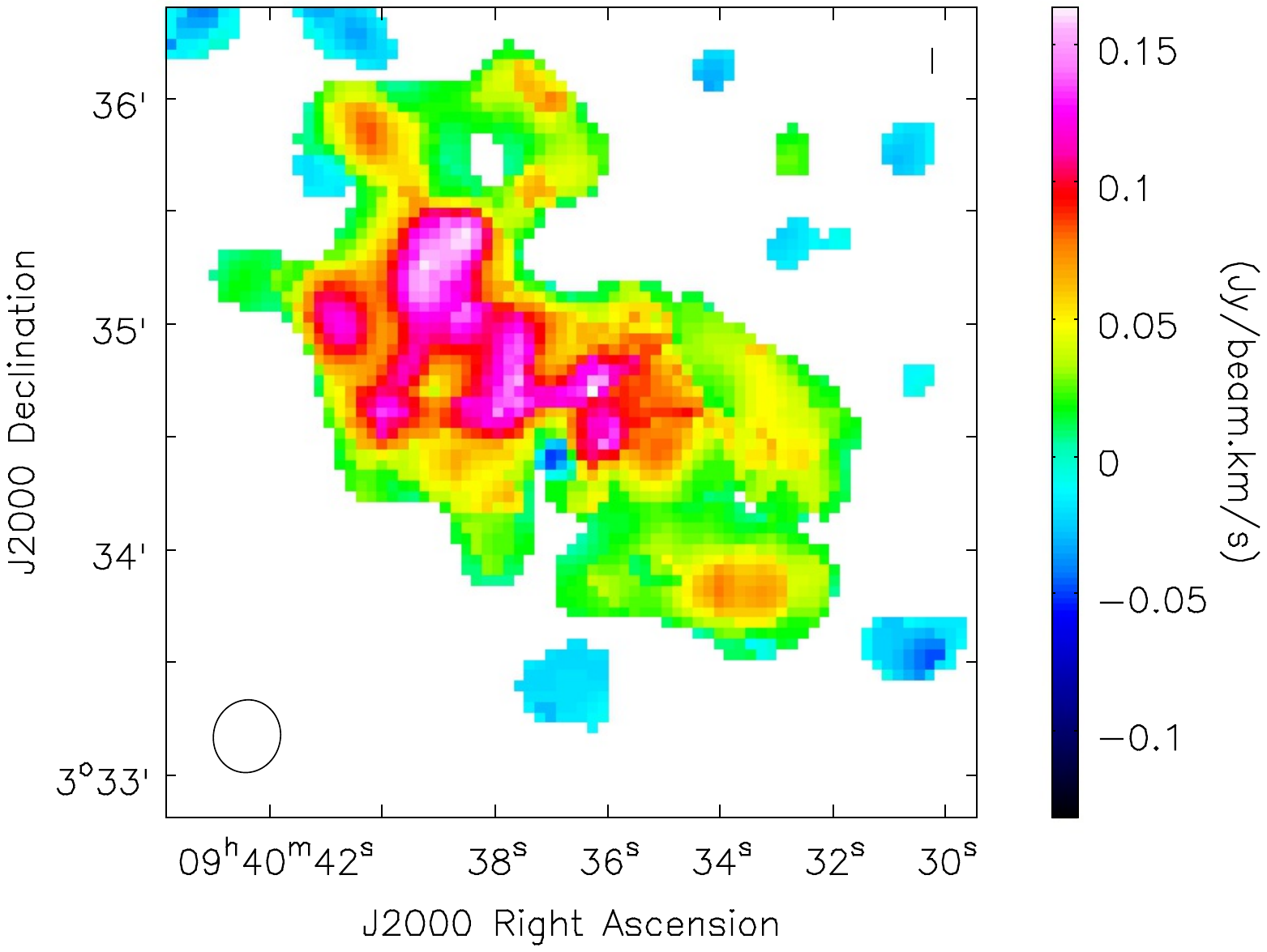}
	\hskip -1mm
	\includegraphics[bb = 25 400 500 750,clip, scale=0.5]{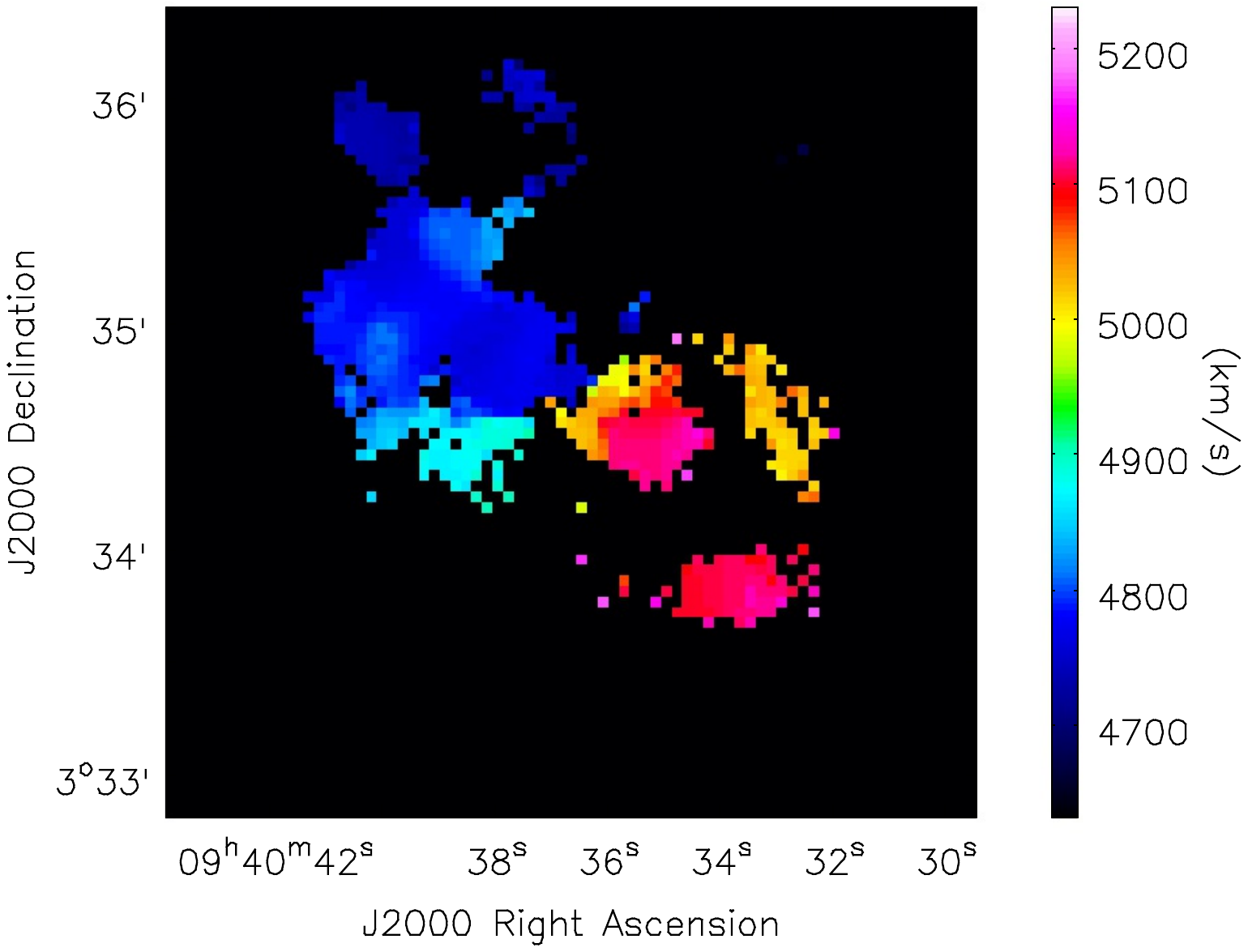}
	}
	}}
      \caption{ Left: The NGC 2960 moment-0 map. The masking is
        described in \S \ref{sec:moment maps}. We mask out the pixels below 2.5 $\sigma$
          in a map that is smoothed spatially and over two velocity
          channels ($40 ~\rm{km~s^{-1}}$).  We then construct the moment-0 map from
          the original data using this mask. The ellipse in the bottom left corner represents the beam.  Right: The NGC
        2960 Gaussian fitted velocity field.}
        \label{fig:ngc2960}
\end{figure*}

\begin{figure*}[h]
	\centerline{
	\vbox{
	\hbox{
	\hskip -0 in
	\includegraphics[bb = 25 400 500 750,clip, scale=0.52]{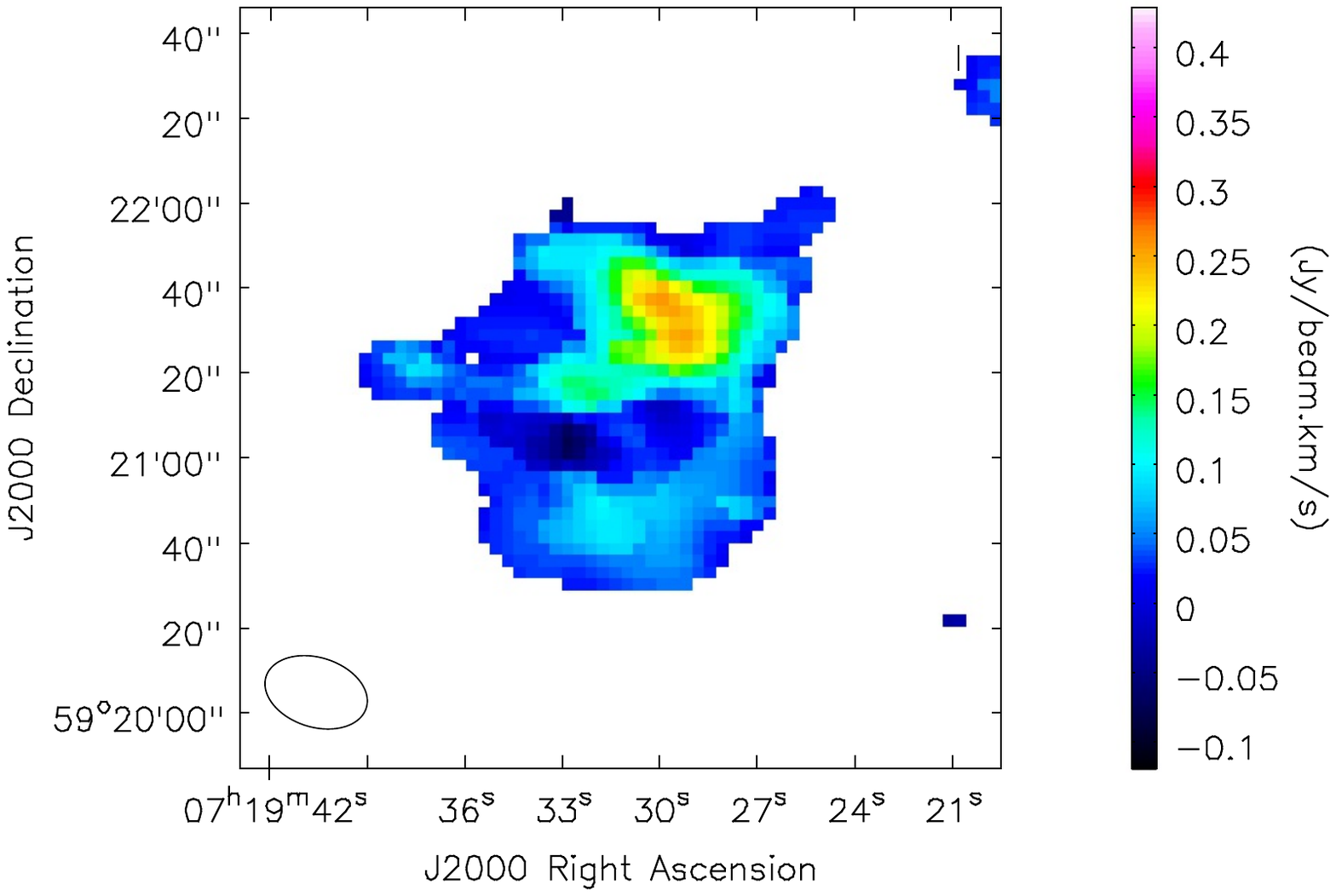}
	\hskip -1mm
	\includegraphics[bb = 25 400 500 750,clip, scale=0.52]{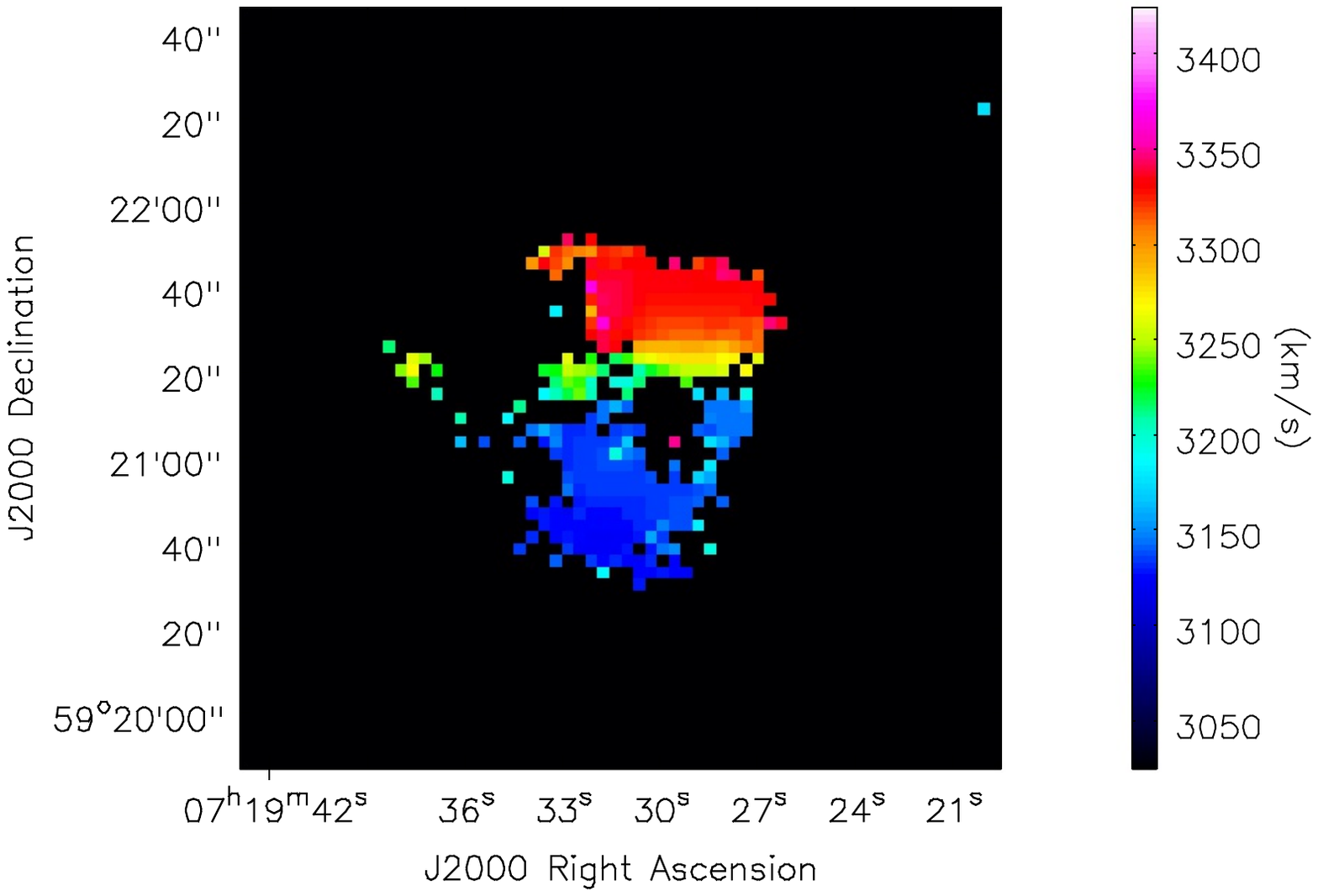}
	}
	}}
	\caption{
          Left: The UGC 3789 moment-0 map. The masking is described in
          \S \ref{sec:moment maps}. We mask out the pixels below 2.5 $\sigma$
          in a map that is smoothed spatially and over two velocity
          channels ($40 ~\rm{km~s^{-1}}$).  We then construct the moment-0 map from
          the original data using this mask. The ellipse in the bottom left corner represents the beam size. Right: The UGC 3789 Gaussian
          fitted velocity field.}
        \label{fig:ugc3789}
\end{figure*}

Non-parametric integrated intensity moment-0 maps of the five galaxies
are shown in the left columns of Figures \ref{fig:ngc2748} to
\ref{fig:ugc3789}. In order to improve the SNR, we used the CASA tool
{\it image.moments}, which masked pixels with no signal in producing
the moment maps. The masks were produced as follows: the data cube was
first smoothed spatially over one beam and spectrally over two image
planes ($20~\rm{km~s^{-1}}$ for NGC 2748 and NGC 7582; $40~\rm{km~s^{-1}}$ for NGC 1194, NGC 
2960, and UGC 3789). 
Then, pixels with smoothed intensity below a
certain threshold were masked. The threshold was set to 3~$\sigma$ for
NGC 2748, NGC 7582, and NGC 1194, and 2.5~$\sigma$ for NGC 2960 and UGC
3789, chosen to optimize the signal seen in the moment-0 maps. 

\subsection{Parametric Fitted Velocity Fields}
\label{sec:fit Vfield}

To extract rotation curves from the \ion{H}{1} data cube, it is
conventional to fit a velocity-field model. The two-dimensional
velocity-field model is parametrized by geometrical parameters and the
rotation curve. Then, the model is fit to the two-dimensional
velocity-field data derived from the 3-D data cube. To construct the
velocity field, one assigns a representative velocity to each
spectrum in the data cube.  In the ideal case of a simple rotating disk
with small random motions, the projected rotational velocity will
correspond to the peak velocity in the spectrum. In practice there are
different ways to measure the velocity at each position, including the
peak velocity, the intensity-weighted mean velocity, and parametrized
fitting \citep[e.g., Gaussian fitting,][]{deBlok08}.  We seek the
velocity measure that can best recover the true projected rotation
velocity in the case of finite resolution, sensitivity, disk warping
and overlapping velocities along the line of sight.

\begin{figure*}[h]
	\centerline{
	\vbox{
	\hbox{
	\includegraphics[ scale=0.45]{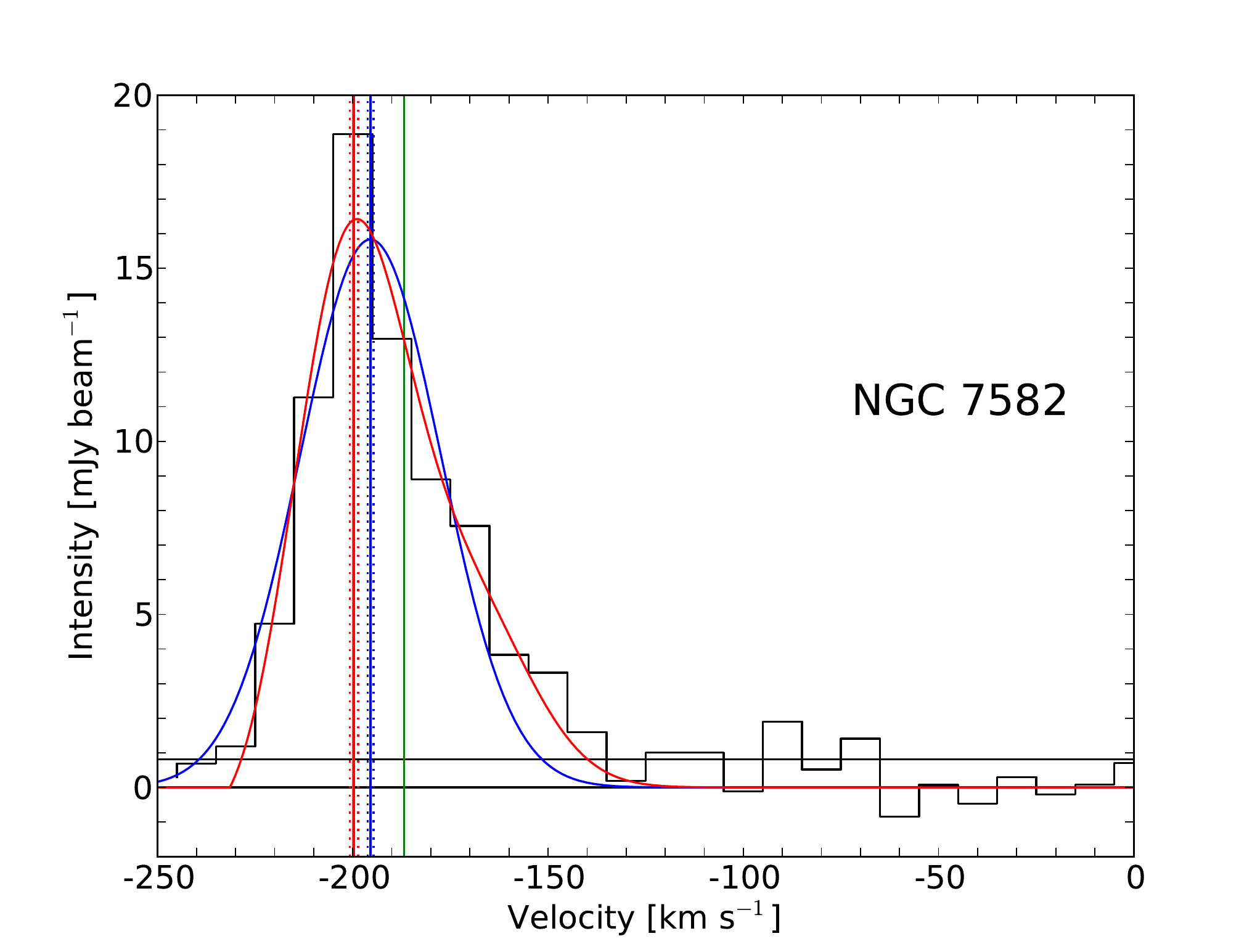}
	\hskip -0.4 in
	\includegraphics[ scale=0.45]{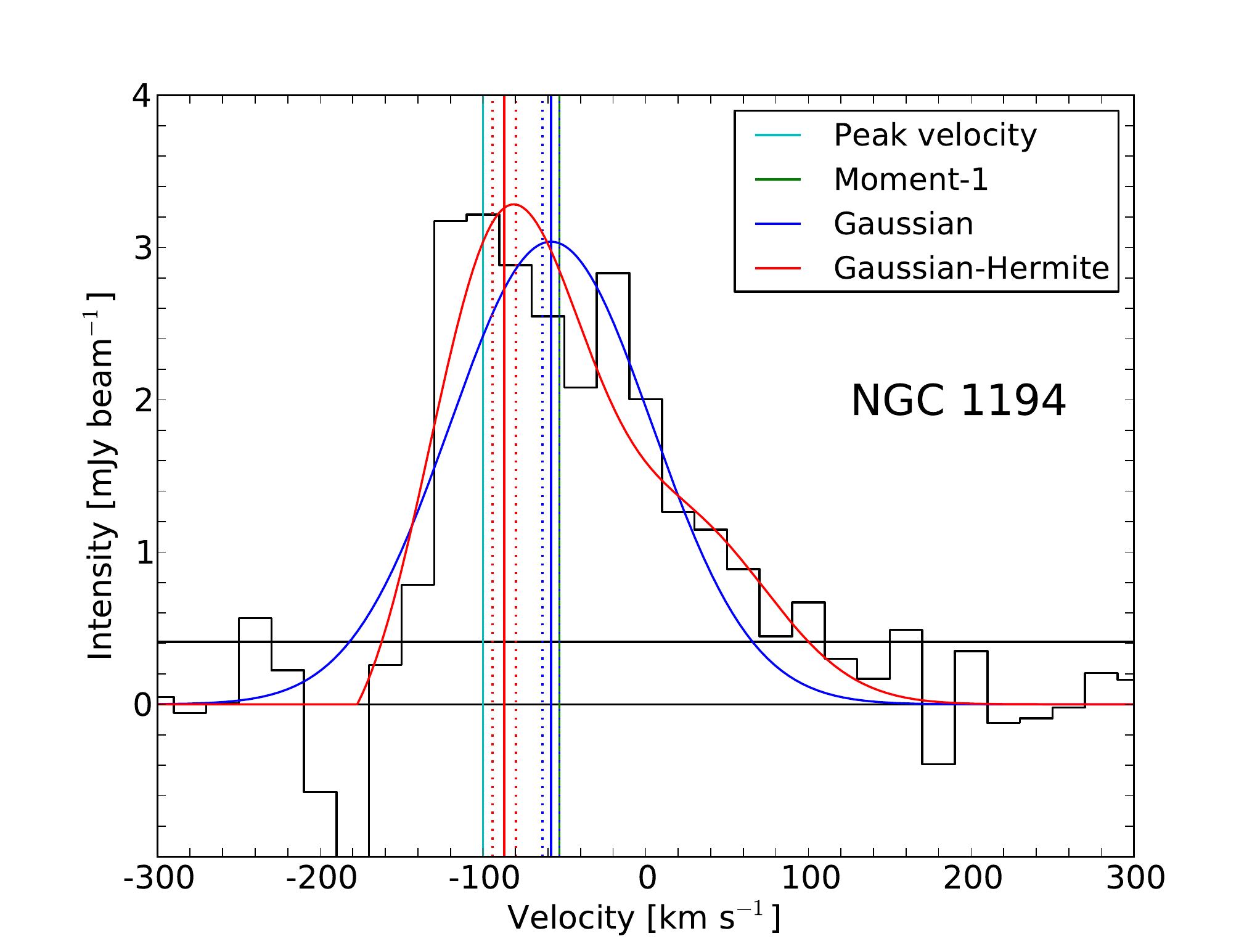}
	}
	}}
        \caption{
          Comparison of different velocity assignments. The cyan vertical
          line is for peak velocity, green for moment-1 velocity, blue
          for the Gaussian fit, and red for the Gauss-Hermite fit. Dashed vertical
          lines are the fitting errors. Black
          horizontal lines show the zero and one sigma
          intensities. The spectrum on the left is from the NGC
            7582 \HI data cube, while the one on the right is from the
            NGC 1194 data cube. }
        \label{fig:compare_spec_step}
\end{figure*}

Measuring the peak velocity is the most straightforward approach, but the
measurement is sensitive to the noise. It may pick up a noise spike
instead of the real signal when the SNR is low. Moreover, it is
discretized to the velocity channel width, which is 10 or 20
km~s$^{-1}$ in our case. The intensity-weighted mean (moment-1)
velocity is less sensitive to the noise, but will be biased if the
spectrum is not symmetric. For example, beam smearing will produce a
wing towards the systemic velocity, causing the intensity-weighted
mean velocity to be skewed towards lower values (see the left panel
of Fig. ~\ref{fig:compare_spec_step}).

Another alternative is parametrized fitting, such as with a Gaussian
line profile. The Gaussian fit picks up the intensity peak if the line
profile is symmetric, and it is not as sensitive to the noise as is the
peak velocity, so Gaussian fitting is our preferred approach.
However, at high SNR if the line shape is asymmetric, then the mean
velocity of the best-fit Gaussian profile will be biased toward the
wing (see the right panel of Figure ~\ref{fig:compare_spec_step}). In
this case, we can use a Gauss-Hermite expansion to capture the
non-Guassian line shape \citep{vanderMarel93}. For example, the
skewness asymmetry is taken into account using the third order
Hermite polynomial $H_3(x)$
\begin{align}
x :&= \frac{v-\bar{v}}{\sigma}, \nonumber \\
f(x) &= \frac{\gamma}{\sqrt{2\pi }\sigma} \exp \left[-\frac{1}{2}\left(x\right)^2 \right][1+h_3 
\mathrm{H_3}(x)], \nonumber \\
\mathrm{H_3}(x) &= \frac{1}{\sqrt{6}}(2\sqrt{2}x^3-3\sqrt{2}x).
\end{align} 
Here the skewness parameter $h_3$ quantifies the asymmetry of the line
shape. The intensity-weighted mean velocity $v_{m, 3}$ of this line
shape $f(v)$ is calculated by \citet{vanderMarel93} to be $v_{m, 3} =
\bar{v} + \sqrt{3} \sigma h_3$ to first order. Compared to a Gaussian,
the mean velocity $v_{m, 3}$ measured by Gauss-Hermite fitting is less
biased towards the wing and closer to the peak, while compared to the
peak velocity the Gauss-Hermite fit is also less sensitive to the
noise. Differences in the velocity fields derived from
  these different methods and their effects on the resulting rotation
  curves are discussed in Appendix \ref{sec:app_RC}. We find that for
  high SNR data, the rotation curve derived from the moment-1 velocity
  field is systematically lower than from the Gaussian/Gauss-Hermite
  velocity field at the $\sim$15$\%$ level, and that the peak-velocity
  produces noisy rotation curves for low SNR data. 

Considering the advantages of Gaussian and Gauss-Hermite fitting, we
utilized these two methods to construct the velocity field. 
In order to minimize contamination from noise, we
applied the noise masking described in \S\ref{sec:moment maps}.  We
then performed a Markov Chain Monte Carlo (MCMC) fitting of a Gaussian
line shape to the five \ion{H}{1} data cubes. We assume the
  likelihood function of intensity is Gaussian, and the sigma of this Gaussian is
a constant throughout the data cube.  The sigma was measured from the RMS
intensity of line-free channels in the cube.

For the three highest-SNR galaxies, i.e., NGC 2748, NGC 7582,
  and NGC 1194 (SNR $>$ 14), there were regions along the major axis
with high reduced-$\chi^2$ values, i.e., where the
  cumulative distribution function of $\chi^2$ is larger than 95
  percent, indicating deviations of the line shape from a Gaussian,
mostly because of the line asymmetry described above. To better
capture the skewness of the line, we fitted these high
$\chi^2$ spectra with a third-order Gauss-Hermite function. The
final velocity fields of NGC 2748, NGC 7582, and NGC 1194 were
constructed using a mixture of Gaussian and Gauss-Hermite fitting. The
lower SNR cubes of NGC 2960 and UGC 3789 (SNR $\sim$ 6) do
  not have enough SNR to show deviations from single Gaussians. 

Higher-order deviations from the Gauss-Hermite line shape, such as the
fourth-order kurtosis ($h_4$), still exist in smaller particular regions after 
using the third-order polynomial; along the major axis of the edge-on galaxy 
NGC 1194 the line-of-sight overlap is severe, and along the spiral
arms in NGC 7582 the gas kinematics are complex. We did not apply
higher-order fits as the inclusion of the $\rm{H_4}$ polynomial would not
change the mean velocity to the first order \citep{vanderMarel93}.

One advantage of the MCMC fitting is that the uncertainty in each
fitted parameter can be estimated from the standard deviation of the
probability distribution probed by the chain. The uncertainties in the
velocities are especially useful. A large velocity uncertainty may
indicate that the spectrum contains no emission line, only noise. We
therefore performed a further masking to exclude pixels with velocity
uncertainty larger than 80 km~s$^{-1}$. The final velocity fields
are shown in the right columns of Figures~\ref{fig:ngc2748}
to~\ref{fig:ugc3789}.

\subsection{Rotation Curves}
\label{sec:fit RC}

Rotation curves are derived by fitting the velocity fields (\S
\ref{sec:fit Vfield}) with a rotating disk model.  This model consists
of coplanar concentric rings, each with its own rotation velocity. The
width of each ring is chosen to be 14\arcsec, corresponding to the
typical beam size. We do not use tilted-ring modeling, but we
  show in Appendix \ref{sec:app_RC} that the rotation curves derived
  with a tilted ring analysis are consistent with those presented
  here.

The rotation-curve fitting was performed with the same MCMC method. We
  assume that the likelihood function of velocity is Gaussian, and use the
velocity uncertainty estimated in \S \ref{sec:fit Vfield}, such that
noisier pixels with larger velocity uncertainties are down-weighted
naturally. We fit five geometrical parameters, the center $x_0$,
$y_0$, the systemic velocity $v_{\rm{sys}}$, the inclination $i$, and
the position angle P.A., together with the rotation velocity, at each
ring. These parameters describe the velocity-field model. For the
megamaser galaxies (NGC 1194, NGC 2960, and UGC 3789) we fix the
central positions ($x_0$,~$y_0$) to the VLBI maser positions, and we
fit the kinematic center of NGC 2748 and NGC 7582 as free parameters.
The $v_{\rm{sys}}$, $i$, P.A., and rotation velocities are fitted as free
parameters with the exceptions of the inclination in the case of NGC 2960
and UGC 3789, and the P.A. of UGC 3789, where these values cannot be
constrained by the data.  Therefore, we fix these three values to
the HyperLeda values \citep{Paturel03}, derived from the optical
images. The best-fit geometrical parameters are listed in Table
~\ref{tab:best fits}.

\begin{figure*}[ht]
        \centering
	\includegraphics[bb = 30 0 800 500,clip, scale=0.60]{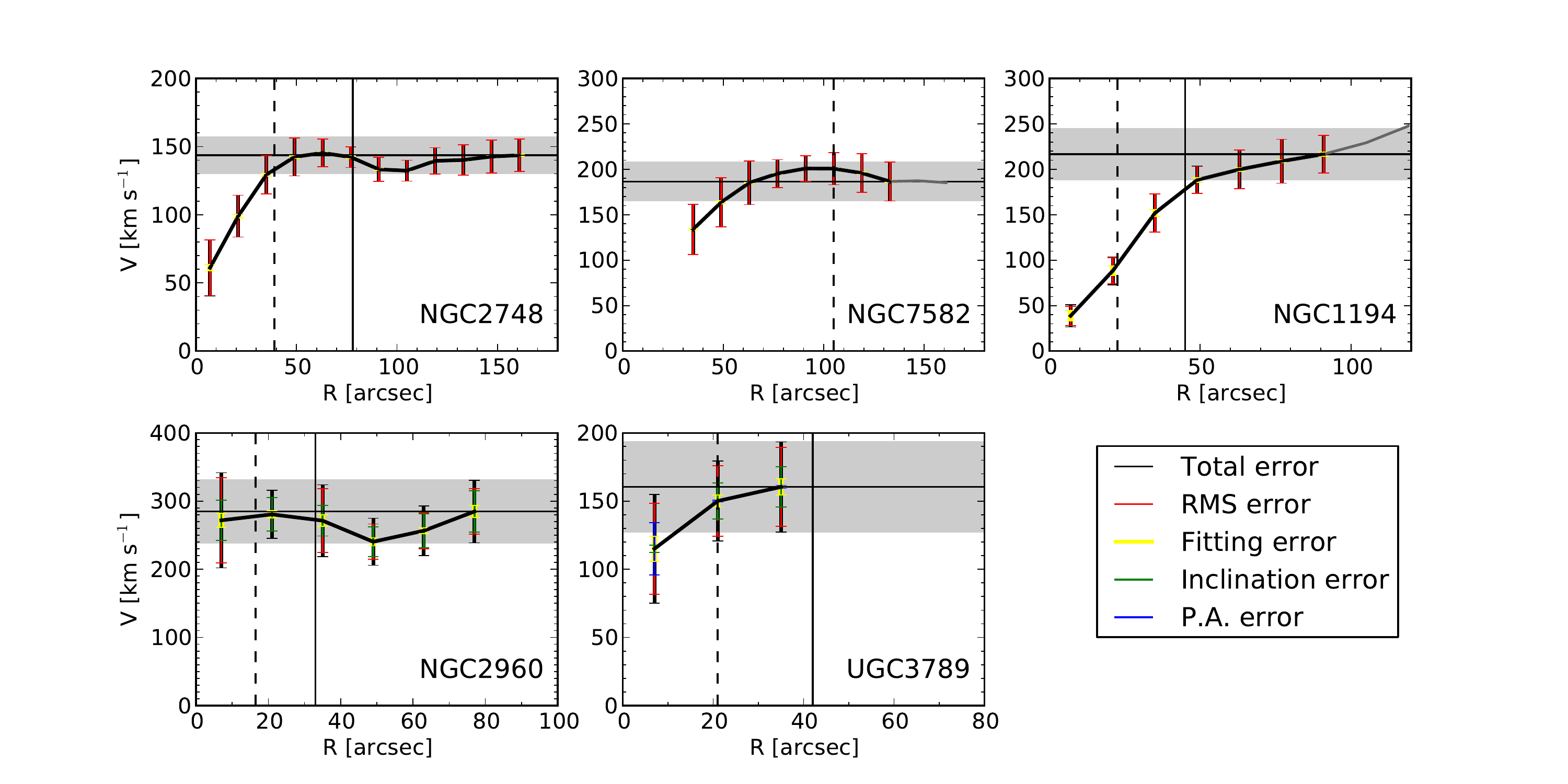}
        \caption{
          Rotation curves of the five galaxies. Red error bars are the
          RMS errors, which dominate the error budget in all
          cases. Yellow is the MCMC fitting
          error. Green/blue is the error contribution from the
          inclination/position angle uncertainty. Black error bars are
          the total error, calculated by the quadratic sum of all the
          error sources. The circular velocity $\Vc$ as measured from the
          outer-most radial bin is denoted with the black horizontal lines with the errors 
          marked with the grey shaded region. This $\Vc$ error includes the observational error and the 
          rotation curve variation. The $R_{\rm{25}}$ radii are marked as the solid 
          black vertical lines and dashed black horizontal lines are the $R_{\rm{25}}/2$. 
          All of the five galaxies have rotation curves extending beyond $R_{\rm{25}}/2$, 
          and NGC 2748, and NGC 1194, and NGC 2960 are beyond $R_{\rm{25}}$. 
          For NGC 7582 and NGC 1194 the last two bins, linked by the dark gray lines, are noisy bins and thus not used for the $\Vc$ measurements.}
        \label{fig:RC_Final}
\end{figure*}

The best-fit rotation curves are plotted in Figure \ref{fig:RC_Final}.
 Various error sources in the rotation curves are considered,
  including the fitting errors estimated by the standard deviation in
  the Markov chains, the RMS errors estimated by the RMS variation of the residuals in a ring, and the errors induced by
  uncertainties in inclination and P.A.  It is commonly found that
  the formal fitting errors in the rotation curve fits are unrealistically small
  \citep[e.g.,][]{deBlok08}, likely because the rotating disk model
  fails to capture features in the observed velocity field such that
  the residuals are non-Gaussian, causing the fitting
  errors to appear artificially small (5-20 times smaller than the RMS
  error). As shown in Appendix
  \ref{sec:app_RC}, the RMS error is larger than or comparable to the
  differences between various velocity assignment methods and different 
  parameterized rotating disk models, and therefore provides 
  a conservative estimate
  of the potential systematic uncertainties. For NGC 2960 and UGC 3789, we also
include the uncertainties in the literature inclination and/or
P.A. values. The uncertainties contributed by these values are estimated by
recalculating the rotation velocities assuming an inclination and/or
P.A. within $\pm 5\degree$ of the HyperLeda value. The final
error is the quadratic addition of the considered error sources.

\section{$\mvc$ Relation}
\label{sec:MBH-Vc}

\subsection{The Sample}
\label{sec:5.1}

Our primary sample is an updated list of galaxies with both dynamical
$\mbh$ and spatially resolved $\Vc$ measurements, as tabulated in
Table \ref{tab:VcReliability} and \ref{tab:MBHVc}. There have been a
number of compilations of $\mbh$ and $\Vc$ \citep{Kormendy11,
  Beifiori12}; we introduce a few significant improvements over
these. First, we update five maser and three stellar/gas dynamical
$\mbh$ measurements. In addition to our five VLA \HI rotation curve
$\Vc$ measurements, there are a few other $\Vc$ measurements included
in this sample from the literature that are not compiled in previous
$\mvc$ studies. We also reexamine the literature $\Vc$ values with a
careful reliability analysis, to ensure that our measurements and the
literature values are derived in a consistent manner. Thus, our $\Vc$
values may differ from previous work. In addition to this primary
sample, we also consider a secondary sample composed of galaxies with
three spatially unresolved $\Vc$ (single dish \HI measurements) and
fifteen black hole mass upper limits. This secondary sample is
described in Appendix \ref{sec:app_sec} and plotted in
Fig. \ref{fig:MBHVc} for comparison, but is not used in our scaling
relation fitting analysis.

\subsubsection{Circular Velocities $\Vc$}
\label{sec:Vcsample}

Table \ref{tab:VcReliability} lists the $\Vc$ values, sources, and
rotation-curve properties of our primary sample.  We examined each
rotation curve from the original literature to assign a $\Vc$ value.
Because $\Vc$ is used as a tracer of the dark matter halo potential in
the outer parts of the galaxy, we assign $\Vc$ as the rotation
velocity at the largest observed radius $R_{\rm{o}}$, which is consistent
with the definition in \citet{Ferrarese02}. If the inclination
correction is applied in the original literature, we use this
inclination-corrected rotational velocity, otherwise we apply a simple
$1/\sin(i)$ inclination correction using the optically determined
inclination from HyperLeda \citep{Paturel03}. The uncertainty
  in $\Vc$ that results from our inclination correction treatment is
  estimated to be less than $\pm 10 \%$ for a typical
  inclination error of $\pm 5\degree$ in the optically derived inclination
  \citep[e.g.,][]{Dressler83} at an average inclination of $\sim
  60\degree$. This is comparable to or smaller than the $\Vc$
measurement errors described below. 

The observational error in $\Vc$, listed first, is taken from the
original literature.  If it is not presented then we assume a $10 \%$
observational error, which is typical among our sample. The second
error stands for the uncertainties due to rotation curve variation. In
the ideal case where the rotation curve is flat at large radius, there
is one uniquely defined $\Vc$. However, if the rotation curve is not
flat but keeps rising or starts to fall, then $\Vc$ depends on where
the outermost observed radius $R_{\rm{o}}$ is located. To take this
uncertainty into account, we assign a second error equivalent to the
amplitude of variation in the rotation curve.  While these error
assignments may overestimate the real error between the observed $\Vc$
and the true halo potential, we hope to avoid overestimating the
intrinsic scatter $\epsilon_{int}$ in the $\mvc$ relation. The final
error used in the $\mvc$ relation fit is taken as the larger of the
observational and rotation curve variation error, as listed in Table
\ref{tab:MBHVc}.

There are two galaxies, NGC 3998 and NGC 5128 (Centaurus A) in our
primary sample that have no rotation curve measurements but have
spatially resolved \HI data. For NGC 3998 we adopt the $\Vc$ value
and observational error estimated by the original literature
\citep{Knapp85}. For NGC 5128, we determine $\Vc$ using the published
P-V diagram, and apply an inclination correction using the inclination from
HyperLeda \citep{Paturel03}. To reflect our lack of knowledge about
the rotation curve trends, we assign conservative rotation curve
variation errors of $20 \%$ in these two cases.

How faithfully $\Vc$ reflects the potential of the dark matter halo
depends on whether the rotation curve is probing the halo-dominated
part of the galaxy. Therefore we compare the outermost observed radius
of the rotation curve $R_{\rm{o}}$ with the optically determined
$R_{25}$, which is the $B = 25$ mag arcsec$^{-2}$ isophote from RC2
\citep{RC2} shown in Column (6) of Table \ref{tab:VcReliability}. It
is worth noting that if $R_{\rm{o}}/R_{25}$ is small, as in the case
of many optical rotation curves, then the rotation curve may be
dominated by baryons rather than the dark matter halo, in turn biasing our 
measurement of intrinsic scatter in the $\mvc$ relation.
  Thus, we divide galaxies into four groups according to the spatial
  extent and shape of their rotation curves. The first
  group has the largest spatial extent ($R_{\rm{o}} > R_{25}$),
  followed by the second group ($R_{25} > R_{\rm{o}} > R_{25}/2$). The
  third and fourth groups both have short rotation curves ($R_{\rm{o}} <
  R_{25}/2$), but in the third group the rotation curves flatten 
  while those ones in the fourth group are still rising
  until the last bin. We discuss the reliability of the $\Vc$
  measurements for these groups in \S \ref{sec:fit_results} and will
  conclude that only the first two groups ($R_{\rm{o}} > R_{25}/2$) have
  reliable $\Vc$ for our $\mvc$ analysis.  

\subsubsection{The $\mvc$ and $\msigma$ Samples}
\label{sec:MBH}

Table \ref{tab:MBHVc} lists $\Vc$, $\mbh$, $\sigmastar$ and
other galaxy quantities. The black hole mass $\mbh$, stellar velocity
dispersion $\sigmastar$, morphology, and distance are adopted from
\citet{McConnell12}, except for NGC 4526 \citep{Davis13}.
The method of $\mbh$ measurement is also listed. 
 We note potential caveats in the $\mbh$
measurements of three galaxies as pointed out by \citet{McConnell12}. 
\citet{Lodato03} measured the black hole mass in
NGC 1068 and found a non-Keplerian maser disk. For NGC 2748,
\citet{Atkinson05} cautions that the determination of the disk center
may be affected by heavy extinction. Finally, the black hole mass in NGC 7457
as measured by \citet{Gebhardt03} could be overestimated if the central
bright source excluded in their dynamical modeling is a nuclear
cluster instead of an AGN. While we keep these three galaxies in our sample, 
we note these points of caution.
  The $\sigmastar$ that we adopt from \citet{McConnell12} is the light-weighted 
  stellar velocity dispersion within one effective
  radius. However, for some galaxies the $\sigmastar$ is lower than
  in previous studies \citep[e.g.,][]{Gueltekin09, Beifiori12}, as the BH
  sphere of influence is excluded in the velocity integration to avoid
  contamination. NGC 3998 and NGC 4594 in our primary sample, as well
  as NGC 1399 and NGC 4486 in our secondary sample, have $\sigmastar$
  updated by \citet{McConnell12} with this correction. 
  We also note that there is a new distance update from megamaser
  measurements for UGC 3789 \citep{Reid13}, but we adopt the
  distance in \citet{McConnell12}, which is consistent with
  \citet{Reid13}.

\subsection{Fitting Method}

To quantitatively analyze the $\mvc$ and $\msigma$ relations, we fit
each scaling relation with a power-law functional form
\begin{align}
  \log~(\mbh/\msun) &= \alpha+\beta\log~(\Vc/200 ~\rm{km~s^{-1}}), \\
  \log~(\mbh/\msun) &= \alpha+\beta\log~(\sqrt{2}\sigmastar/200 ~\rm{km~s^{-1}}).
\end{align}
Here $\alpha$ is the intercept and $\beta$ is the logarithmic slope of the relation. 
 We use $\sqrt{2}\sigmastar$ to compare with $\Vc$,
  motivated by the widely used singular isothermal sphere model where
  $\Vc = \sqrt{2}\sigmastar$ \citep{Binney&Tremaine}. Observationally
  \citet{Ferrarese02} also found an almost linear relation between
  $\Vc$ and $\sigma$ with a ratio close to $\sqrt{2}$.

We use a maximum-likelihood fitting method similar to
\citet{Gueltekin09} that takes intrinsic (or cosmic) scatter
$\epsilon_{int}$ into account, with modifications regarding the $\Vc$
error treatment.  For simplicity we assume that the probability
distribution of both the observational errors and intrinsic scatter are 
normal in log space, and all the observational errors are symmetrized in log
space before fitting. Here we denote
\begin{align}
  \label{}
  \mu_i &= \log~( M_{\rm{BH}, {\it i}}/\msun), \\
  \nu_i &= \log~( V_{\rm{c}, {\it i}} /200 ~\rm{km~s^{-1}}) \\
  &\rm{or} ~ \log~(\sqrt{2}\sigma_{*, {\it i}} /200 ~\rm{km~s^{-1}}),
\end{align}
and the normalized Gaussian error distribution to be
 $G_{\epsilon}(x)$ with mean zero and standard deviation $\epsilon$. 

To take the error in the independent variable ($\Vc$ and $\sigmastar$ in
this case) into account, we have an error term $\sigma_\nu$ in the
likelihood function. We write the likelihood for observing one galaxy
($\mu_i$, $\nu_i$) given an intrinsic scaling relation $\mu = \alpha
+\beta \nu$ with intrinsic scatter $\epsilon_{int}$ to be
\begin{align}
  \label{eq:Li_Guassian}
  l_{i} = G_{\sqrt{\sigma_{\mu}^{2} + \beta^{2}  \sigma_{\nu}^{2} + \epsilon_{int}^{2}}}(\mu_{i} - \alpha - \beta 
\nu_{i}).
\end{align}
This treatment of the errors in the independent variable is less
computationally expensive than the Monte Carlo method used in
\citet{Gueltekin09}. 
We then maximize the total likelihood
\begin{align}
  \label{}
  	\mathcal{L} = \prod_i l_i
\end{align}
to get the best-fit scaling relation
 parameters, the intercept $\alpha$, the slope $\beta$, and 
the intrinsic scatter $\epsilon_{int}$.
 This maximum-likelihood method
is similar to the minimum $\chi$-square method described in
\citet{Tremaine02}, except for the $1/{\epsilon \sqrt{2\pi}}$ normalization factor of the Gaussian function.  We adopt the $\Delta \chi ^2 =1$ confidence limit
as an error estimate for the fitted parameters $\alpha$, $\beta$, and
$\epsilon_{int}$.

\subsection{Fitting Results}
\label{sec:fit_results}

We find that $\mbh$ is correlated with both $\Vc$ and \sigmastar, with
Spearman rank correlations of $\rho=0.62$ ($p$-value
$2\times10^{-3}$) and $\rho=0.67$ ($6\times10^{-4}$) respectively, for
the reliable sample of 22 $R_{\rm{o}} > R_{25}/2$ galaxies discussed
below, meaning that a correlation exists for both of the
  relations. Using this sample, the best-fit $\mvc$ relation is
$\alpha = 7.43^{+0.13}_{-0.13}$, $\beta = 3.68^{+1.23}_{-1.20}$, and
$\epsilon_{int} = 0.51^{+0.11}_{-0.09}$, and is plotted as blue lines
in Figure \ref{fig:MBHVc}. There is no significant correlation between
the residuals in the $\mvc$ relation and $\sigmastar$ (Spearman
$\rho=0.33$, $p$-value$=0.13$).  The parameters of different $\mvc$
samples as well as the comparison with the $\msigma$ relation are
shown in Table \ref{tab:M_BH-Vc_results}.

To determine a reliable sample and investigate how the spatial
extension of the rotation curves affect the reliability of the $\Vc$
measurements, we derive the scaling relations using sample with
different $R_{\rm{o}}$ criteria. The first group of fourteen
  galaxies ($R_{\rm{o}} > R_{25}$) should be the most reliable sample
with rotation curves that probe the outer, halo-dominated, region of
the galaxy.  The second sample of eight galaxies ($ R_{25} >
  R_{\rm{o}} > R_{25}/2$) also gives the same intrinsic scatter as
the first group, showing that the $\Vc$ measurements with $ R_{25} >
R_{\rm{o}} > R_{25}/2$ are as reliable as the first group for our
purpose. When we include groups three and four, the eleven shorter rotation curves 
 with $R_{\rm{o}} < R_{25}/2$,
which are predominantly measured in the optical, 
the intrinsic scatter increases significantly. It is a sign that for
these short rotation curves the halo potential is not represented by
the observed $\Vc$, and the intrinsic scatter will be inflated
artificially if they are included in the analysis. We therefore rely
only on the $R_{\rm{o}} > R_{25}/2$ sample of twenty-two galaxies
  from the first two groups for our main scientific discussion. 
 
Using this $R_{\rm{o}} > R_{25}/2$ sample, the $\mvc$ relation has an
intercept consistent with the $\msigma$ relation, if $\Vc$ corresponds
to $\sqrt{2}\sigmastar$. While the slopes of the $\mvc$ and $\msigma$ relations
are consistent within our sample, our $\mvc$ relation slope
falls on the lower end of reported slopes from studies of the
$\msigma$ relation using larger samples, \citep[e.g.,][]{Gueltekin09,
  McConnell12}, possibly due to our limited dynamic range. The
intrinsic scatter in the $\mvc$ relation of $\epsilon_{int}=
0.51^{+0.11}_{-0.09}$ is similar to that in the $\msigma$ relation of
$\epsilon_{int} = 0.48^{+0.10}_{-0.08}$ when using matched $R_{\rm{o}}
> R_{25}/2$ samples, but is larger than the $\msigma$ relation scatter
$\epsilon_{int} = 0.38$ for the entire sample in \citet{McConnell12},
which includes both early and late-type galaxies. The implications of
our fits are discussed in \S \ref{sec:discussion}.

Some elliptical galaxies also have $\Vc$ measured from dynamical
modeling. Although these measurements may involve different
systematics than the rotation-curve-derived $\Vc$ for disk galaxies,
we plot six such ellipticals from \citet{Beifiori12} in Fig. \ref{fig:MBHVc} as an attempt to
probe the $\mvc$ trend at the high mass end. NGC 1399, NGC 3379, NGC
4374, NGC 4472, and NGC 4486 (M87) have $\Vc$ measured by \citet{Kronawitter00} using stellar
kinematics, while NGC 3608 is measured by \citet{Coccato09}.
Note that for NGC 4486 (M87), \citet{Kronawitter00} measured $\Vc = 508 \pm 38$ \kms, while \citet{Murphy11} measured a much larger  $\Vc = 800^{+75}_{-25}$ \kms, highlighting the challenges in this method. All the black hole masses are
from \citet{McConnell12}. When these ellipticals are added, the slopes
of the $\mvc$ relation and $\msigma$ relation increase significantly
(by $\sim 1 ~\sigma$ to $\beta=5.03^{+0.69}_{-0.75}$ and
$\beta=4.77^{+0.57}_{0.58}$ respectively). \citet{McConnell12} found a
similar steepening in the $\msigma$ relation when considering both
early and late-type galaxies. On the other hand, the intrinsic scatter
decreases slightly by about $0.5 ~\sigma$ to $\epsilon_{int} =
0.46^{+0.10}_{-0.09}$ and $\epsilon_{int} = 0.45^{+0.08}_{-0.07}$ for
the $\mvc$ and $\msigma$ relations respectively. These results will
not be used in our scientific discussion, as their interpretation
awaits better understanding of the correspondence between the $\Vc$
from the dynamical modeling and from \HI rotation curves.

\subsection{Comparison with the Literature}

\citet{Volonteri11} re-analyze the 25 galaxies in \citet{Kormendy11},
which have dynamical $\mbh$ measurements and $\Vc$ from spatially
resolved \HI or stellar kinematics, using the methods of
\citet{Gueltekin09}. They constrain the $\mvc$ relation to have
$\alpha = 7.39 \pm 0.14$, $\beta = 4.33 \pm 0.93$, and $\epsilon_{int}
= 0.53 \pm 0.10$, consistent with our results.

\citet{Beifiori12} study the $\mvc$ relation using 28 galaxies with
$\mbh$ compiled by \citet{Gueltekin09}, and $\Vc$ from resolved \HI
kinematics, unresolved \HI line width, or dynamical models of stellar
kinematics in early type galaxies. Their $\mvc$ relation parameters
are $\alpha = 7.82 \pm 0.15$, $\beta = 3.29 \pm
0.61$, and intrinsic scatter $\epsilon_{int} = 0.51 \pm
0.09$. Their intercept $\alpha$ is different from our result, possibly
because more elliptical galaxies are included in their sample. Their fit 
for the slope has a larger dynamic range because they include more 
elliptical galaxies, but is still consistent with ours, as is their intrinsic
scatter measurement.

\section{Possible Systemic Uncertainties}
 
Intrinsic scatter in the scaling relations provides an important
discriminant between different BH scaling relation scenarios.  Various
observational errors in $\Vc$ have been taken into account in \S
\ref{sec:MBH-Vc} to avoid contaminating the intrinsic scatter
measurement, including the inclination correction, the 
observational error, rotation curve variations, and the uncertainties
due to short rotation curves. Here we discuss whether other systematic 
uncertainties, such as sample bias, may affect the interpretation of,
and the comparison between, the $\mvc$ and $\msigma$
relations. Also, in order to interpret the $\mbh-\mdm$ relation from
the observed $\mvc$ relation, we discuss the scatter that may be
introduced in the $\Vc - \mdm$ conversion.  We conclude first that the
comparison between the $\mvc$ and the $\msigma$ relations should be
fair even in the face of selection effects, and 
second, that the scatter introduced by the $\Vc-\mdm$
conversion or halo triaxiality is not important compared to other sources of 
uncertainty.

\subsection{Selection Effects}
\label{subsec:Vc_unceratinty}

If our sample is subject to selection effects, it may not represent
the true distribution of objects. Selection effects may exist for both
BH and circular velocity measurements. First, it has been argued that
analysis excluding $\mbh$ upper limits might be biased toward more
massive BHs at a given velocity dispersion, especially for late-type
galaxies, since only massive BHs have a sphere of influence large
enough to be spatially resolved
\citep{Gueltekin11}. \citet{Beifiori12} includes a large number of
$\mbh$ upper limits in their fits to the $\mvc$ and $\msigma$
relations, which should be closer to the true distribution. They find
a slope $\beta$ for the $\mvc$ relation consistent with ours, but
cannot derive a reliable intrinsic scatter.  While concerns over 
$\mbh$ selection biases cannot be excluded definitively, we use the
same sample for both $\msigma$ and $\mvc$ relations,
such that if $\mbh$ biases exist, they have the same effect on the
two relations.

 In addition to selecting against low-mass BHs, we
  also select against galaxies that have no $\Vc$ from \HI rotation
  curves, which includes massive elliptical (gas-poor) galaxies, and
  disk galaxies that either have only optical (mostly covering a limited radial distance) or 
  no rotation-curve measurements. With no stellar dynamical $\Vc$
measurements for ellipticals, our dynamic range is limited compared to
the \msigma\ relation in other studies, \citep[e.g.,][]{Gueltekin09,
  McConnell12}. While selection biases in $\Vc$
  measurements also exist, at least there is no obvious bias on the
  intrinsic scatter inherent with these selections. 

\subsection{Uncertainties in Translating $\Vc$ to $\mdm$}
\label{subsec:conversion_uncertainties}

If we wish to interpret the $\mvc$ relation as a reflection of the
underlying $\mbh-\mdm$ relation, then scatter may be introduced in
the conversion from $\Vc$ to $\mdm$. To estimate the magnitude of this scatter, we take
the Navarro-Frenk-White \citep[NFW;][]{NFW} model as an
example. This model is a mass profile that is parametrized by 
the inner density $\rho_{\rm{s}}$ and the characteristic
inner radius $r_{\rm{s}}$, or equivalently the virial mass $M_\mathrm{vir}$
and the concentration $c_\mathrm{vir} \equiv R_\mathrm{vir}/r_s$. The
virial radius $R_\mathrm{vir}$ is defined as the radius within which
the averaged halo density exceeds the background density by a certain
factor, the virial mass $M_\mathrm{vir}$ is the mass enclosed within
$R_\mathrm{vir}$, and the virial velocity $V_\mathrm{vir}$ is the
rotation velocity at $R_\mathrm{vir}$. Because there is a one-to-one
correspondence between $M_\mathrm{vir}$ and $V_\mathrm{vir}$, the main
uncertainty actually comes from the conversion between
$V_\mathrm{vir}$ at the virial radius (typically a few hundred kpc)
and $\Vc$ at the observed radius 
\citep[typically 10-20 kpc,][]{Ferrarese02}. The observed $\Vc$ is close to the maximum
velocity $V_\mathrm{max}$ occurring at radius $r_\mathrm{max} \sim
2.16 ~r_{\rm{s}}$, which is a few to a few tens of kpc. The ratio
$V_\mathrm{max}/V_\mathrm{vir}$ depends on $c_\mathrm{vir}$
\begin{align}
\frac{V_\mathrm{max}}{V_\mathrm{vir}} = 0.46\sqrt{\frac{c_\mathrm{vir}}{\ln(1+c_\mathrm{vir})-c_
\mathrm{vir}/(1+c_\mathrm{vir})}}. 
\end{align} 
The scatter in concentration is estimated by \citet{Bullock01} and
corrected by \citet{Wechsler02} to be $\Delta \log_{10} c_\mathrm{vir}
= 0.14$ dex, which translates to an error of 0.035 dex in
$V_\mathrm{max}$ or 0.14 dex in $\mbh$ for a slope of $\beta=4$ in the
$\mvc$ relation.  There is also a weak dependence of
concentration on the virial mass, which affects the slope of the
$\mvc$ relation in principle.

Additionally, if the halo is not spherically symmetric, as assumed by
the NFW profile, but is prolate or triaxial, there will be extra
uncertainty in the $\Vc - \mdm$ conversion. \citet{franx92} estimate
that the low observed scatter in the Tully-Fisher relation constrains the
ellipticity of the halo potential in the disk plane to be $\leq 0.1$.
Triaxiality can at most contribute a scatter of 0.026 dex in $\Vc$ or
0.1 dex in $\mbh$ in the $\mvc$ relation.

Removing these additional sources of uncertainty in quadrature,
assuming that they are uncorrelated, changes the final intrinsic
scatter very little (0.48 rather than 0.51). Thus we conclude that
those sources of scatter are small compare to outstanding measurement
uncertainty.

\section{Discussion}
\label{sec:discussion}

We will now consider the implications of our new fits for the
evolution of BHs and galaxies.  In \S \ref{subsec:FeebackorMerging},
we hope to gain new insight into whether AGN feedback is required to
explain the BH-bulge scaling relations
\citep[e.g.,][]{Silk98,Hopkinsetal06}, or whether galaxy merging alone
may lead to tight BH-bulge correlations in massive galaxies
\citep[e.g.,][]{Peng07,Jahnke11}. In \S \ref{sec:upper-limits} we
discuss the implications of $\mbh$ upper limits as outliers in the
$\mvc$ relation.

\subsection{Is It Feedback or Merging That Sets the Scaling Relations?}
\label{subsec:FeebackorMerging}

We address the relative importance of feedback and merging empirically
by asking which can account for features in the observed $\mvc$ and
$\msigma$ relations.  We start by reviewing the expected slope,
outlier behavior, and intrinsic scatter of the scaling relations in
the context of each scenario. We will find that
  while more theoretical guidance and a larger sample size are needed to
  make inferences from the outliers and slope measurements, the intrinsic scatter
  is already an available and useful discriminant between the two
  scenarios. 

In the galaxy merging scenario \citep{Peng07, Hirschmannetal10, Jahnke11}, the
correlations between $\mbh$ and galactic quantities (e.g., $M_{*}$,
$M_\mathrm{bulge}$, $M_\mathrm{DM}$) emerge from a large number of
mergers simply by the central limit theorem, even without a physical
mechanism linking the two.
In the context of this scenario, we expect the following features of
the scaling relations. First, as the number of mergers increases and
the galaxies grow larger, the scatter between $\mbh$ and galaxy
properties will decrease and the distribution will converge to a
tighter correlation. Therefore, it is expected that the intrinsic
scatter $\epsilon_\mathrm{int}$ will decrease as the number of mergers
increases towards larger $\Vc$ and $\sigmastar$. Outliers, with BH
masses deviating from a power-law scaling relation,
become increasingly unlikely towards higher masses.
Second, the logarithmic slope $\beta$ of the correlation between two
linear quantities, e.g., $\mbh$ and $M_\mathrm{DM}$, should be close
to unity if the impact of growth via accretion for any component is
small. In other words, the average ratio $\left< M_\mathrm{BH} /
  M_\mathrm{DM} \right>$ should be similar across all masses. If we
translate this slope to the $\mvc$ relation, assuming $M_\mathrm{DM}
\propto V_\mathrm{c}^{3~\mathrm{-}~3.32}$ \citep{Ferrarese02}, we
should find the slope of the $\mvc$ relation, $\beta$, to be close to
$3 - 3.32$, somewhat smaller than, but allowed by, our
observations. Lastly, the $M_{\mathrm{BH}} - M_\mathrm{DM}$ relation
should be tighter than other scaling relations, e.g., the
$M_{\mathrm{BH}} - M_\mathrm{*}$ or $\mgal$ relations, because
$M_\mathrm{DM}$ simply adds during merging and does not depend on the
baryonic physics that $M_*$ and $M_\mathrm{bulge}$ are subject to. For
example, during merging, extra factors of star formation and disk to bulge conversion
have to be considered for the evolution of $M_\mathrm{bulge}$, and this
will make the scatter in the $\mgal$ relation larger than that in the
$M_{\mathrm{BH}} - M_\mathrm{DM}$ relation, if the baryonic regulation mechanism is not present.
However, we note that diffuse DM accretion in principle may also enhance the scatter in $M_\mathrm{DM}$.

Feedback, on the other hand, provides a physical mechanism that
couples BH mass directly with the galaxy potential well
\citep[e.g.,][]{Silk98}. If the BH is massive enough, its energy or momentum
output can expel gas from its
vicinity and quench the further growth of both
the BH and the galaxy. This feedback loop sets an upper limit on the BH
mass for a given potential well depth, measured by the stellar velocity
dispersion $\sigmastar$ of the bulge or $\Vc$ of the dark matter
halo. However, it is unclear whether this self-regulation
process takes place on the scale of the BH sphere of influence, the
bulge \citep{Debuhr10}, or the dark matter halo \citep{DiMatteo03,
  Booth10}. Some predictions of the feedback scenario are as
follows. 
First, the scaling relations form an upper boundary for $\mbh$. The 
BH cannot grow above the relation via accretion (although over-massive BHs 
could already be in place from massive seeds \citet{Volonteri09}).  
Second, the slope of
the $\msigma$ or the $\mvc$ relation is predicted to be five for energy
feedback \citep{Silk98} and four for momentum feedback
\citep{Murrayetal05}. Third, the correlation between
BH mass and the potential indicator ($\sigmastar$ or $\Vc$) on the
feedback scale should be tighter than on other scales.

We here compare the statistics of our $\Vc$ sample with the
predictions discussed above regarding the outlier properties, slope,
and the intrinsic scatter, respectively.
There are no obvious outliers above the $\mvc$
relation in our sample, possibly due to limited sample size. However,
there are bulgeless galaxies with $\mbh$ upper-limits that are low compared to
their $\Vc$, which will be further discussed in \S
\ref{sec:upper-limits}.  Regarding the measured slopes, with such
limited dynamic range our data are still inadequate to discriminate
the two scenarios. Both scenarios are consistent with the wide range
of allowed $\mvc$ slopes ($\beta$ = 2.48 - 4.91).

However, the comparison of intrinsic scatter between the two relations
can provide useful constraints on the origin of the scaling
relations. Our measurements show that the scatter in the $\mvc$
relation is similar to that in the $\msigma$ relation, within the
errors. It is worth noticing that the intrinsic scatter in $\mvc$,
$\epsilon_{\mathrm{int}}=0.51^{+0.11}_{-0.09}$ dex, is a
conservatively low estimate, as we have assigned large errors to the
$\Vc$ measurements to account for various observational uncertainties,
including inclination uncertainties and rotation curve
variations. Even so, we still find a value that is consistent with our
measured $\msigma$ scatter, $\epsilon_{int} = 0.48^{+0.10}_{-0.08}$
dex, and that is not smaller than the $\msigma$ intrinsic scatter
$\epsilon_{int} = 0.46$ for late-type galaxies constrained by
\citet{McConnell12} or $\epsilon_{int} = 0.38$ for the entire sample
in \citet{McConnell12}.

We find no evidence that the $\mvc$ (or $\mbh-\mdm$) relation is
significantly tighter than the $\msigma$ relation, in contradiction
with naive expectations from the pure merging scenario. Therefore,
our result disfavors merging as the only
mechanism that ties together BHs and galaxies.  Note that we use
$\sigmastar$ to trace the baryonic mass in the bulge, while
\citet{Jahnke11} looked at the $\mbh-\mbulge$ relation instead. Unfortunately,
observations of $\mbulge$ for lower-mass galaxies with dynamical
$\mbh$ are scarce, so a direct comparison between $\mbh-M_{\rm{DM}}$
and $\mbh-\mbulge$ is not yet possible. If the real scatter in the
$\mbh-\mbulge$ relation is even smaller than in $\msigma$, our conclusion is
stronger. Even if the scatter in the $\mbh-\mbulge$ relation turns
out to be larger than in the $\msigma$ relation, the fact that the
scatter in the $\mvc$ relation is no smaller than that in the
$\msigma$ relation still requires (baryonic) mechanisms beyond the pure
merging scenario.

On the other hand, if AGN feedback is important in regulating BH
growth, and it acts primarily on the bulge scale \citep{Debuhr10}, 
the smaller intrinsic scatter in the $\msigma$ relation could be explained. 
We speculate that on top of the correlations formed by merging, feedback 
further regulates the BH mass according to the bulge potential $\sigmastar$, 
tightens the $\msigma$ relation (not the $\mvc$ relation), and decreases the
intrinsic scatter in a relative sense. 
Alternatively, other baryonic mechanisms that connect the BH to the bulge, e.g., via feeding of bulge stars onto the accretion disk \citep{Miralda05}, 
or star formation-regulated BH growth \citep{Burkert01}, could also contribute to the tighter
$\msigma$ relation. 

\subsection{$\mbh$ Upper Limits in the Two Scaling Relations}
\label{sec:upper-limits}

$\mbh$ upper limits in bulgeless galaxies have also been used to
differentiate the dependence of black hole mass on galaxy halos or
bulges. \citet{Kormendy11} found that $\mbh$ cannot be uniquely
determined by $\Vc$.  Bulgeless galaxies have low (or zero) \mbh\ 
and low \sigmastar\ but relatively high $\Vc$.  Therefore, \sigmastar\ 
is a better predictor of \mbh\ than is $\Vc$ for these galaxies.

We expand the sample of bulgeless galaxies with $\mbh$ upper limits by
including the measurements from \citet{Neumayer12} in our secondary
sample. This sample is described in Appendix \ref{sec:MBHUL} and listed in Tables
\ref{tab:VcReliability_secondary} and \ref{tab:MBHVc_sec}. As can be
seen from Figure \ref{fig:MBHVc}, there are a few upper limits lying
below the $\mvc$ relation but not the $\msigma$ relation.  With a
larger number of galaxies, we confirm Kormendy's statement that the halo 
mass alone does not determine the black hole mass.

However, it is also possible that these bulgeless galaxies do not host
a massive black hole at all.  The presence or absence of a BH seed
may involve different physical mechanisms than those that couple BHs
to galaxies \citep[e.g.,][]{Volonteri11}.  In any case, the upper
limits strengthen our conclusion that bulge scales matter, likely both
in seeding and in BH growth with cosmic time.

\section{Summary}

We refine the measured $\mvc$ relation and compare it to the $\msigma$
relation to gain insight into the mechanisms that drive BH scaling
relations. We perform VLA observations to measure the circular
velocities $\Vc$ for five galaxies with dynamical $\mbh$. Together
with a thorough literature search, we increase the sample size of
galaxies with both $\mbh$ and $\Vc$ measurements to
thirty-three. Twenty-two of these have $\Vc$ evaluated at large enough
radius ($> R_{25}/2$) to be reliable for our scientific purpose.

With this sample, we constrain the power-law $\mvc$ relation to have
an intercept $\alpha = 7.43^{+0.13}_{-0.13}$, slope $\beta =
3.68^{+1.23}_{-1.20}$, and intrinsic scatter
$\epsilon_{int}=0.51^{+0.11}_{-0.09}$. The intrinsic scatter in the
$\mvc$ relation is not significantly smaller than that in the
$\msigma$ relation, showing that \mbh\ does not correlate better with
dark matter halo mass than with bulge properties. This contradicts naive
expectations from pure merging scenarios. Furthermore, we consider a
number of $\mbh$ upper limits in bulgeless galaxies that lie
significantly below the $\mvc$ relation, suggesting that BH masses are
better predicted by the bulge, not the halo, either via its seeding
or accreting mechanism. We thus suggest that pure merging is not
likely to be the only mechanism that drives the scaling relations.
AGN feedback may also be an essential ingredient to tighten the
correlation between BH mass and bulge properties.

We highlight possible future improvements to this work. First,
modeling of the dark matter halo mass distribution, to separate the
halo mass from the baryons using both rotation curves and light
distributions, can improve the halo mass estimation and better
constrain the $\mbh-\mdm$ relation. Second, $\Vc$ measured by stellar
dynamical modeling for elliptical galaxies, if proven to be comparable
to those from \HI rotation curves, can improve the dynamic range and
better constrain the slope of the relation. Third, quantitative
theoretical predictions for the scaling relations, especially the
intrinsic scatter, from both feedback and pure merging scenarios, will
enable more direct comparisons with the empirical relations.
 Ultimately, a large sample unbiased with respect to
  $\mbh$ and $\Vc$, which awaits next-generation telescopes, would be
  the most ideal data set for this study.

\acknowledgements
We thank the referee for a very thorough report that helped us 
improve this manuscript.
We thank K. G\"{u}ltekin, L. C. Ho, N. McConnell, and K.  Jahnke for helpful
discussions, M. Strauss for constructive suggestions, and J. Kormendy for providing valuable references. We
acknowledge the usage of the HyperLeda database (http://leda.univ-lyon1.fr) and NASA/IPAC Extragalactic Database (http://ned.ipac.caltech.edu).

\begin{appendix}

\section{\HI Properties of Individual Galaxies}
\label{sec:HI_prop}
\begin{figure*}[h]
	\centering
	\includegraphics[scale=0.6]{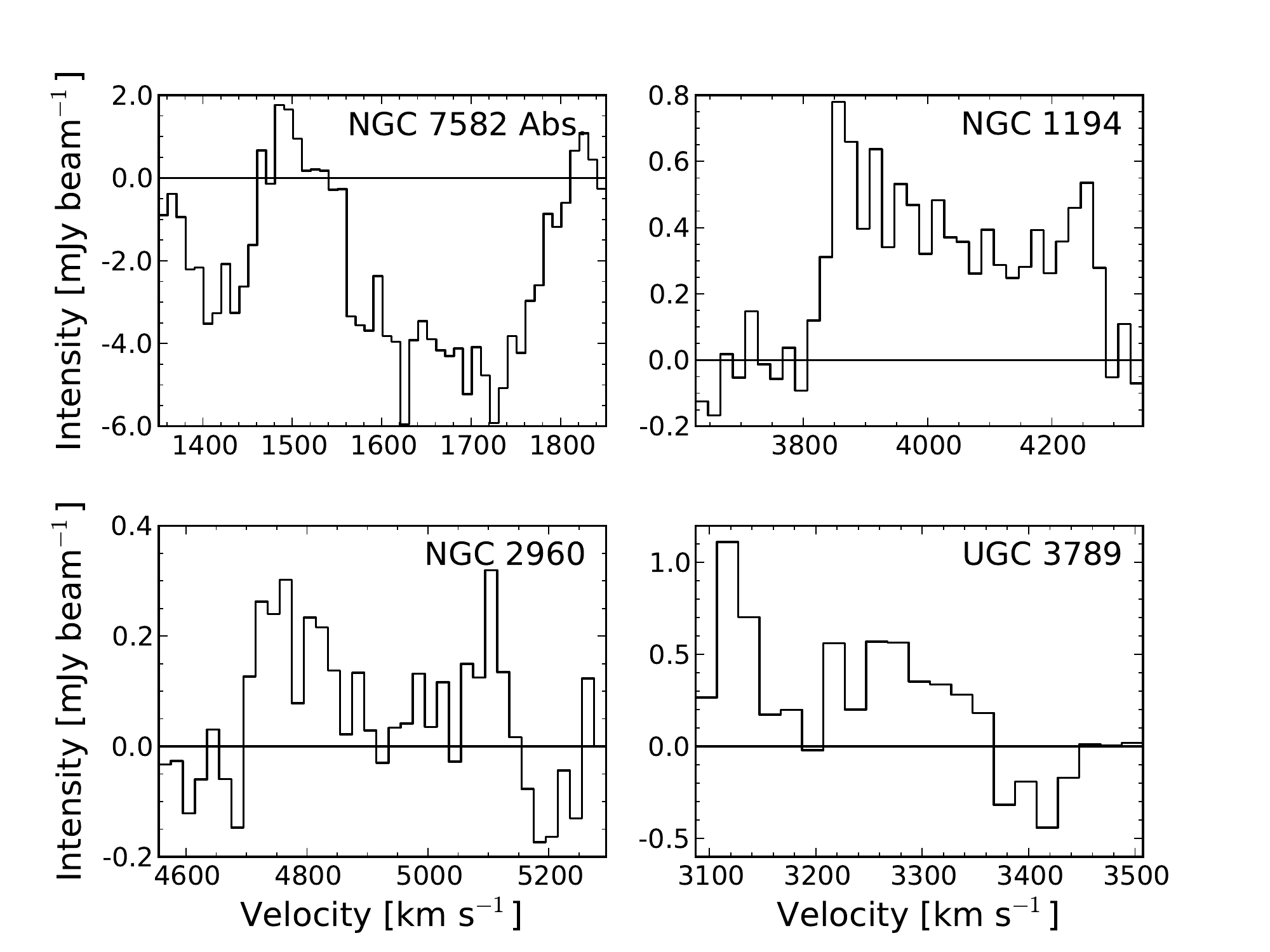}
	\caption{The spectrum of the NGC 7582 central \HI absorption
          feature and the integrated \HI emission spectra of the three lower SNR galaxies (NGC 1194, NGC 2960, and UGC 3789). }
	\label{fig:spec}
\end{figure*}

In addition to the rotation curves kinematics, there is also rich
information about the \HI gas properties in our VLA data cubes. We
discuss the \HI fluxes and masses and the \HI properties of each
individual galaxy in this section. \HI moment maps of these galaxies
are shown in Figures \ref{fig:ngc2748} to \ref{fig:ugc3789} and the
integrated spectra of lower SNR galaxies are shown in Figure \ref{fig:spec}. 

\subsection{\HI Flux and Mass}
\label{sec:HI_mass}

The \HI fluxes and mass are tabulated in Table \ref{tab:HI_properties}.
We measured the \HI flux by integrating the data cube over the
spatial regions as shown in the moment-0 maps. To avoid contamination,
for NGC 1194, we excluded the north-western cloud, which is detached
from the main galaxy as seen from the moment-0 map. We also removed
the central absorption region in NGC 7582.  As an estimate of the \HI
flux error, we also measured the flux from the masked moment-0 maps,
which exclude the noise-dominated channels and therefore can
underestimate the total flux. We take the difference between the two
as the systematic flux error. 
 The \HI masses are calculated using the equation
  \begin{align}
M_{\rm{HI}} = 2.343 \times 10^5 ~\msun ~(1+z) \left( \frac{D_L}{\rm{Mpc}} \right)^2 \left( \frac{\int F_\nu ~dv}{\rm{Jy ~\kms}} \right)
\end{align}
from \citet{Draine11}.  For consistency, we adopt the same distances
as those used for the $\mbh$ measurements in \S \ref{sec:MBH} and Table
\ref{tab:image quality}. Distance uncertainties of $10\%$ are
assumed, and the uncertainties in $M_{\rm{HI}}$ are propagated
from the uncertainties in flux and distance.   
It is common in the literature to 
compare the \HI mass to the total stellar mass, to understand the 
gas fractions and available fuel for future star formation.
Here we use the B-band luminosity ($L_{\rm{B}}$)
as a proxy for the stellar mass and calculate the \HI mass-to-light
ratio ($M_{\rm{HI}}/L_{\rm{B}}$) as a proxy for the H~{\small I}-to-stellar mass ratio.  Using the B-band luminosity from
HyperLeda \citep{Paturel03}, the $M_{\rm{HI}}/L_{\rm{B}}$ ratios of
our galaxies range from $3 \%$ to $22 \%$ (Table
\ref{tab:HI_properties}). This is similar to the range of typical
Sa galaxies, which have an average value $\sim 10 \%$ with a factor of
$\sim 3$ dispersion \citep{Roberts94}, as well as AGN hosts which also
have $M_{\rm{HI}}/L_{\rm{B}} \sim 10 \%$ \citep{Fabello11, Ho08}.
Therefore, our sample galaxies have roughly similar H~{\small I}-to-stellar 
mass ratios as typical disk galaxies and AGN hosts. 

\subsection{NGC 2748}

NGC 2748 is an SAbc galaxy \citep{RC3} at a
  distance of 21.6 $\pm$ 1.4 Mpc.  The \HI observation of this galaxy has the
highest SNR among our five galaxies, with an \ion{H}{1}-based inclination of
$72.6\degree$, compared to $68.1\degree$ from optical data in
HyperLeda \citep{Paturel03}. The fitted \HI systemic velocity is 1482 km~s$^{-1}$. The kinematics show the typical
signature of a rotating disk (Fig. \ref{fig:ngc2748}). The asymmetric
velocity field suggests that slight disk warping may exist. The
angular diameter along the major axis at the 3 $\sigma$ level,
 i.e., where \HI is detected above the 3 $\sigma$ level,
 is
$3\farcm8$. The \HI flux is measured to be $37 \pm 5 ~\rm{Jy~km~s^{-1}}$, which
 corresponds to an \HI mass of $4.2 \pm 0.9 \times 10^9~\msun$.

\subsection{NGC 7582}

NGC 7582 is an SB(s)ab galaxy \citep{RC3} at a distance of 20.6
  $\pm$ 2.4 Mpc. Our observation of this galaxy also has high SNR. 
The inclination derived from the \HI kinematics is $67.9 \degree$,
compared with $68.2\degree$ from HyperLeda \citep{Paturel03}. The
  systemic velocity is 1588 km~s$^{-1}$ as fit by \ion{H}{1}. Its
velocity field shows the clear signature of rotation and is asymmetric
(Fig. \ref{fig:ngc7582}), suggesting a warped disk. The velocity
discontinuity at the edge of the warp coincides with the
location of the spiral arms. Its angular diameter is $3\farcm7$ at the
3~$\sigma$ level. The \HI flux is $20 \pm 3~\rm{Jy~km~s^{-1}}$, which 
corresponds to an \HI mass of $2.0 \pm 0.5 \times 10^9 ~\msun$. 

NGC 7582 shows \HI absorption features against the central continuum
source. The continuum-subtracted spectrum averaged over the central
beam is shown in Figure \ref{fig:spec}. There is an emission peak at a
velocity of 1500 $\rm{km~s^{-1}}$ coinciding in velocity with the
emission in nearby regions. Therefore, we suspect that we are
observing the superposition of a wide absorption feature with narrow
emission. The wide absorption feature has a FWHM of $\sim$ 400
$\rm{km~s^{-1}}$ centered at 1580 $\rm{km~s^{-1}}$, close to the
systemic velocity. Because of contamination from \HI emission, we can
only estimate a lower limit on the absorbed flux. Using the line width
and depth, that flux limit is $> 0.94 ~\rm{Jy~km~s^{-1}}$. The lower
limit on the \HI absorption optical depth is estimated to be $>0.027$.

\begin{figure*}[h]
	\centerline{
	\vbox{
	\hbox{
	\hskip -0.2in
	\includegraphics[bb = 25 400 500 750,clip, scale=0.5]{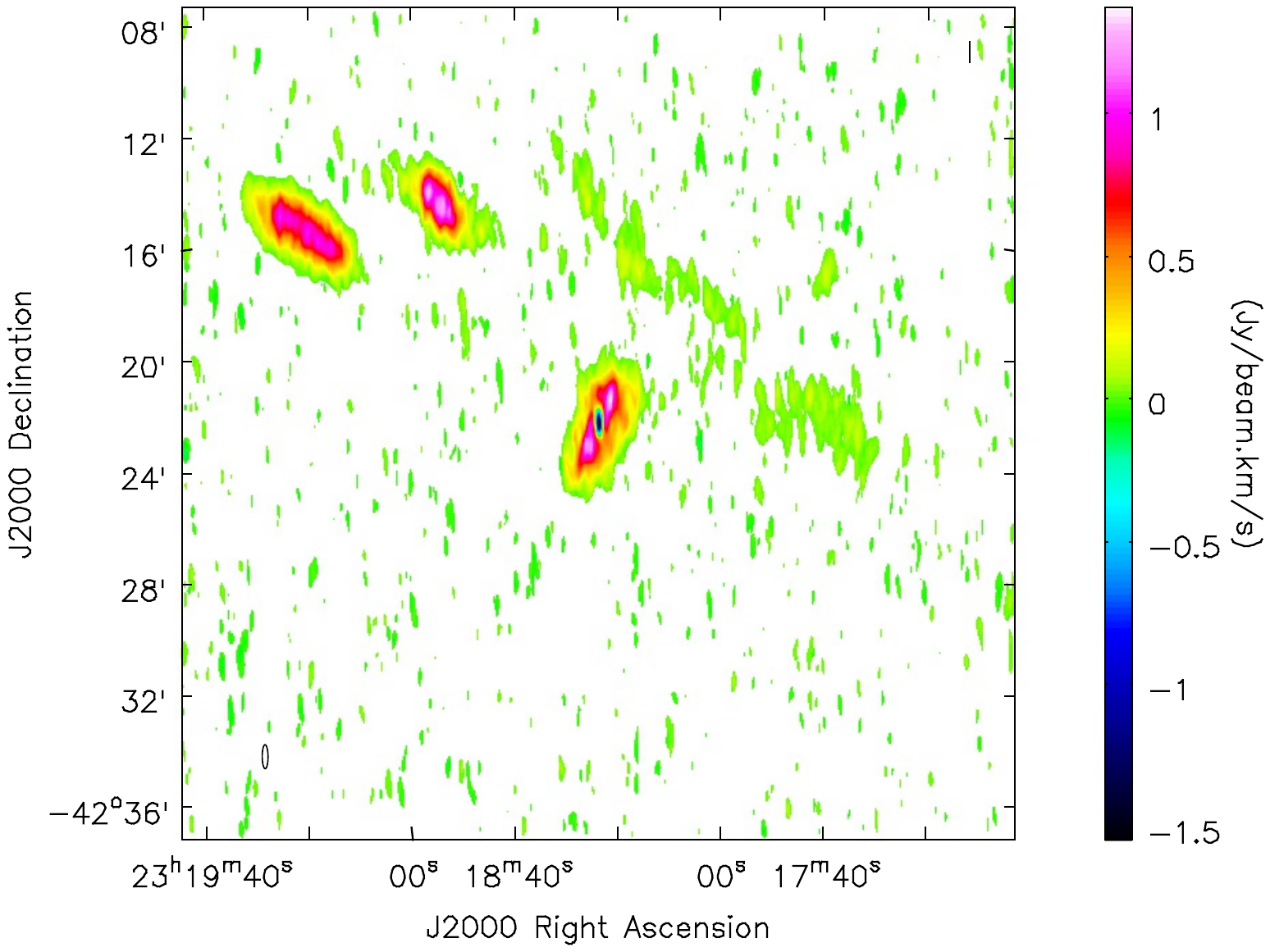}
	\hskip -1mm
	\includegraphics[bb = 25 400 500 750,clip, scale=0.5]{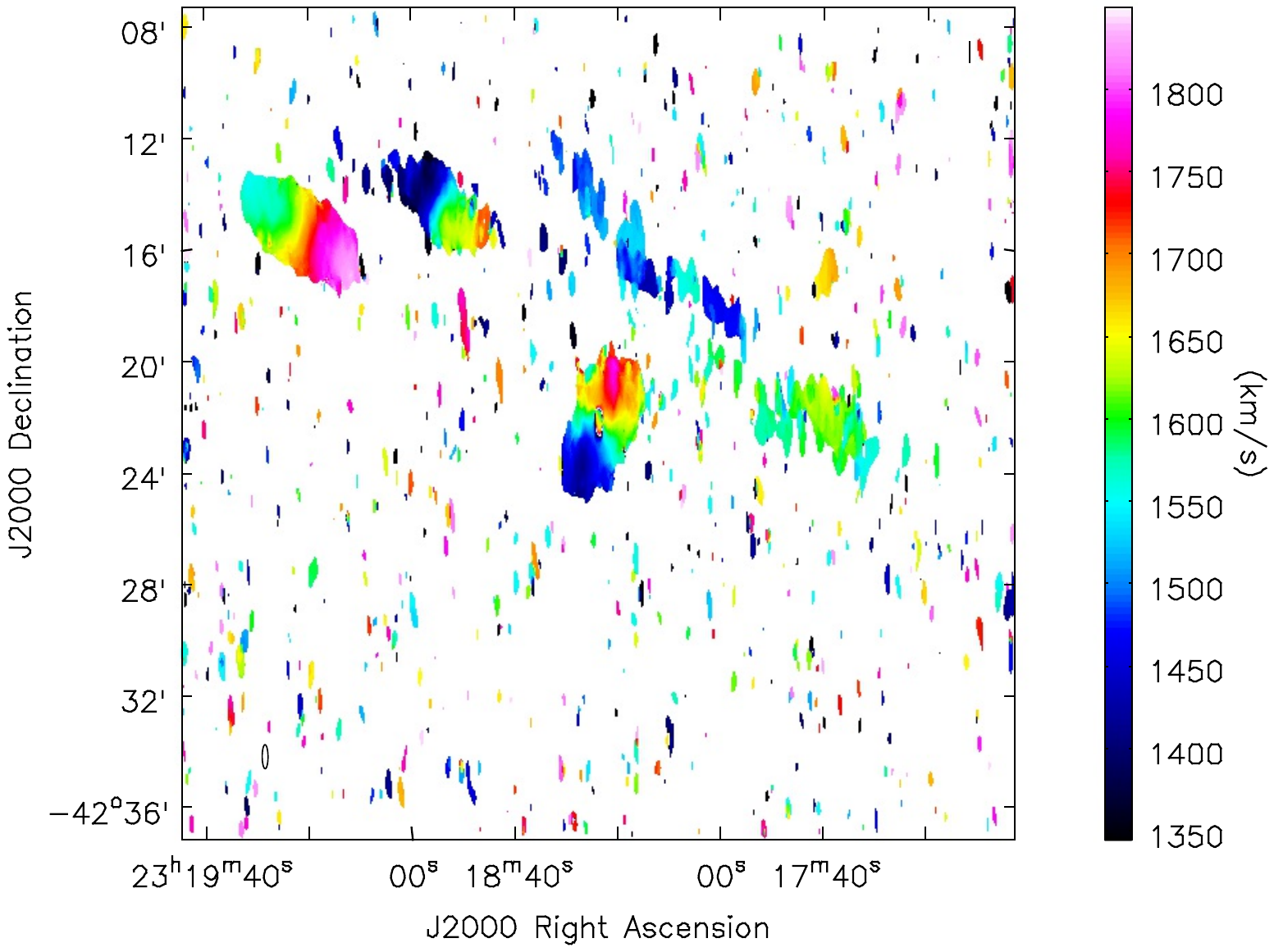}
	}
	}}
	\caption{
          Left: The NGC 7582 moment-0 map on larger scale. The masking
          is the same as in Figure \ref{fig:ngc7582}. We mask out the pixels below 3 $\sigma$
          in a map that is smoothed spatially and over two velocity
          channels ($20 ~\rm{km~s^{-1}}$).  We then construct the moment-0 map from
          the original data using this mask. Two companions,
          NGC 7590 and NGC 7599, and an elongated tidal stream can be
          seen. Right: The moment-1 map of the same field.}
        \label{fig:ngc7582_zout}
\end{figure*}

Finally, there are two \HI companions of NGC 7582 observed in the same
velocity range and sitting $9\arcmin$ and $12\arcmin$ to the north-east respectively. 
These are the companion galaxies NGC 7590 and NGC 7599. There is faint diffuse
\HI emission $\sim 13\arcmin$ long that extends from NGC 7590
to the west of NGC 7582 with a closest distance to NGC 7582 of
  $5\farcm5$. This is possibly tidally stripped gas due to
interactions between these galaxies. Larger scale moment-0 and moment-1 maps
showing these structures are plotted in Figure \ref{fig:ngc7582_zout}.

\subsection{NGC 1194}

\begin{figure*}[h]
	\centerline{
	\vbox{
	\hbox{
	\hskip -0.2in
	\includegraphics[bb = 25 400 500 750,clip, scale=0.5]{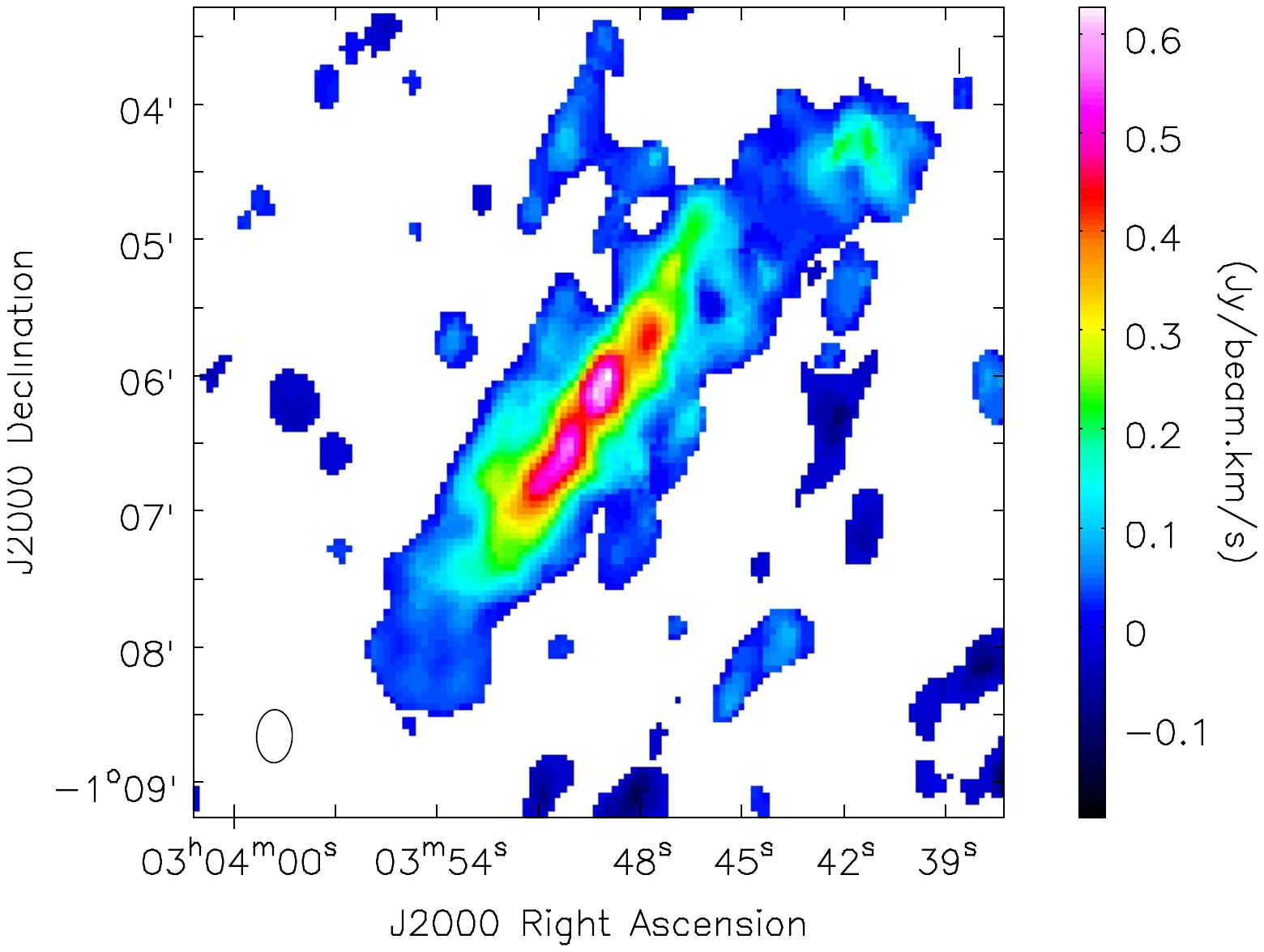}
	\hskip -1mm
	\includegraphics[bb = 25 400 500 750,clip, scale=0.5]{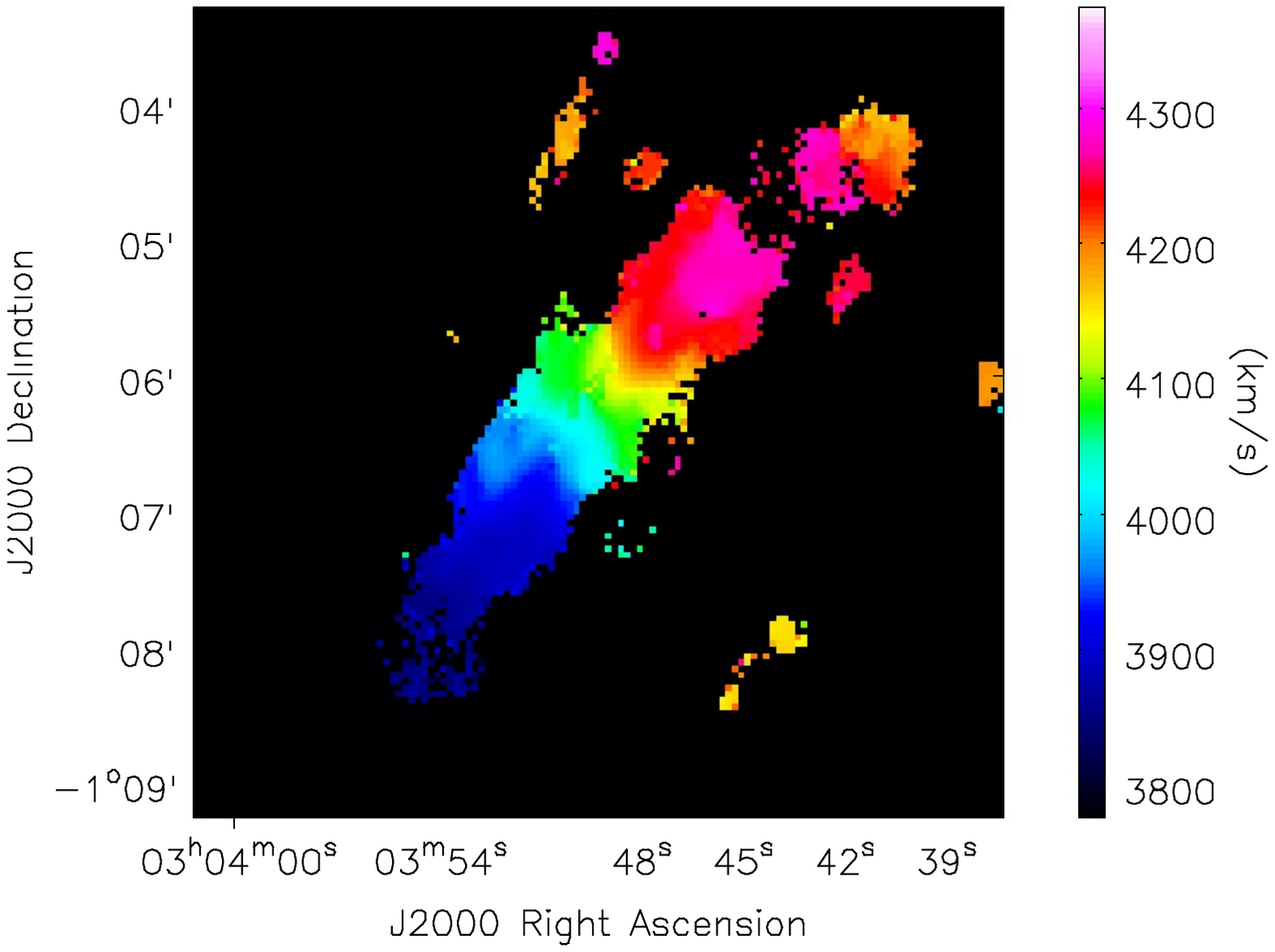}
	}
	}}
	\caption{
          Left: The NGC 1194 moment-0 map. The masking is described in
          \S \ref{sec:moment maps}. We mask out the pixels below 2.5 $\sigma$
          in a map that is smoothed spatially and over two velocity
          channels ($40 ~\rm{km~s^{-1}}$).  We then construct the moment-0 map from
          the original data using this mask. 
          The ellipse in the bottom left corner represents the 
          beam size. Right: The NGC 1194
          Gaussian/Gauss-Hermite fitted velocity field. }
        \label{fig:ngc1194}
\end{figure*}

NGC 1194 is an SA0 galaxy \citep{RC3} at a distance of
53.2 $\pm$ 3.7 Mpc. It is one of our water megamaser galaxies
\citep{Kuo11}. The \HI SNR is sufficiently high to determine an
inclination of $69.1 \degree$, compared to $71.1 \degree$ from
HyperLeda \citep{Paturel03}. The systemic velocity as measured from
the \HI map is 4075 km~s$^{-1}$. The \HI moment-0 map shows an
elongated morphology with angular diameter $2\farcm8$ at the
$3~\sigma$ level, showing that the \HI gas is organized within the
galaxy disk. The velocity field shows a clear velocity gradient due to
rotation (Fig. \ref{fig:ngc1194}). The integrated spectrum
(Fig. \ref{fig:spec}) also has a clear double peaked rotation
signature with the width consistent with our $\Vc$ measurement. The
\HI flux is measured to be $7 \pm 1 ~\rm{Jy~km~s^{-1}}$, corresponding
 to a mass of $5 \pm 1 \times 10^9 ~\msun$
. There is one \HI cloud to the north-west
side of the galaxy that is detached from the main galaxy body (see
Fig. ~\ref{fig:ngc1194}). This cloud has a mass of $5\pm 2 \times10^8 ~M_{\odot}$.

\subsection{NGC 2960}

NGC 2960, also called Mrk 1419, is an Sa \citep{RC3} galaxy at 72.2
$\pm$ 5.1 Mpc and also a water megamaser galaxy from \citet{Kuo11}.
The systemic velocity from \HI is 4939 km~s$^{-1}$. Our map of NGC
2960 has lower SNR (SNR $=6$). Although \HI emission is only seen in
discrete patches of the galaxy, an overall velocity gradient is
observed (Fig \ref{fig:ngc2960}). The detected diameter is $2\farcm5$
at the $1~\sigma$ level from the moment-0 map. The inclination cannot
be constrained by the \HI data alone, and we derive the rotation curve
by adopting the optical inclination from HyperLeda \citep{Paturel03}
of $41.5 \degree$. The integrated spectrum (Fig. \ref{fig:spec}) shows
a double peaked rotation signature and a width that is consistent with
our $\Vc$ measurement. The \HI flux is measured to be $2.2 \pm
0.3~\rm{Jy~km~s^{-1}}$, corresponding
 to a mass of $2.7 \pm 0.5 \times10^9 \msun$.
Previous D-array VLA observations \citep{Kuo08} are
consistent with the spatial extent and velocity from our data, 
but their \HI flux of $1.7\pm 0.3~\rm{Jy~km~s^{-1}}$ is lower than ours. 
 This is understandable as their flux was calculated from the moment-0 map and 
should be lower than from the data cube.

\subsection{Jet in NGC 2960}
\label{sec:ngc2960_jet}

\begin{figure*}[h]
	\centerline{
	\vbox{
	\hbox{
	\hskip -0.2in
	\includegraphics[bb = 25 400 500 750,clip, scale=0.53]{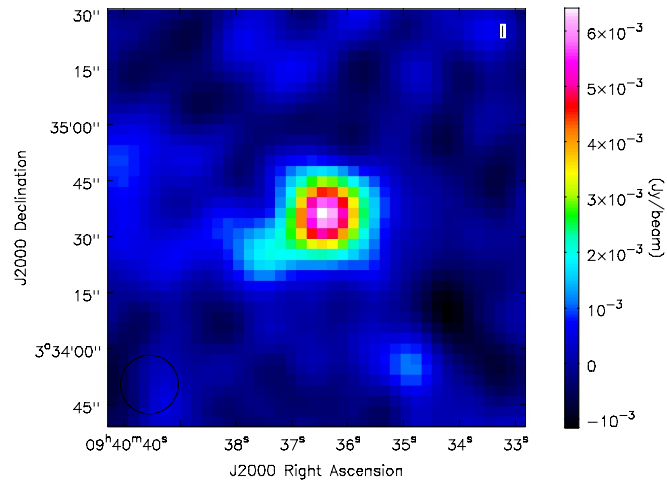}
	}
	}}
	\caption{
          The 20 cm continuum image of NGC 2960. There is extended
          emission on the south-west side of the central point
          source at the water maser position, suggesting that there is
          a jet launched from the galaxy nucleus (\S
          \ref{sec:ngc2960_jet})}
        \label{fig:ngc2960_cont}
\end{figure*}

Our \HI data also provides images of the radio continuum at 20
cm. We find that the radio continuum of NGC 2960 is slightly
extended, suggesting that there is a radio jet launched from the
central black hole (Fig. \ref{fig:ngc2960_cont}). The continuum image
is made from line-free channels on both sides of the \HI line with a
total velocity range of 300 $\rm{km ~s^{-1}}$, and is cleaned to the
$5~\sigma$ level using $robust= 0.5$ weighting. We measure the size
and position angle of the jet using two different methods. First, we
fit a central point source ($15\farcs7$) and subtract it.  We detect
residual emission to the south-east.  Taking into account the
positional uncertainty due to the finite beamsize, this extended
emission is $20 \pm 3 \arcsec$ away from the center at a position
angle of $125 \degree$. The flux in the extended emission is $1.6 \pm
0.2$ mJy.  To estimate the size and P.A. error we also try a second
method to quantify the elongated structure. We fit the whole continuum
image with a two-dimensional Gaussian while allowing the semi-major
and semi-minor axis, central position, and P.A. to vary, without
deconvolution of the beam. The fitted Gaussian has a major axis of $22
\pm 4\arcsec$, and a minor axis of $18 \pm 4\arcsec$ with P.A. $119
\pm 7 \degree$, consistent with the first method. Therefore we
conclude that the jet size is $20\pm5\arcsec$ at a P.A. of $125 \pm
10\degree$. As in all megamaser galaxies studied to date, the 
jet axis is coincident with the rotation axis of the maser disk 
\citep{Greene13}.

\subsection{UGC 3789}

UGC 3789 also hosts a water maser disk, is an SA(r)ab
galaxy \citep{RC3}, and is at a distance of 49.9 $\pm$ 7.0 Mpc
  \citep{Braatz10}. The \HI systemic velocity is 3229
  km~s$^{-1}$. The \HI observation also has low SNR $= 6$ and shows a
ring-like structure with a diameter of $1\farcm3$ at the $1~\sigma$
level on the moment-0 map (Fig. \ref{fig:ugc3789}). There is a
clear velocity gradient. Neither inclination nor position angle
can be constrained from the \HI data alone. The optical inclination $i
= 43.2 \degree$ and position angle $\rm{P.A.} = 164.7\degree$
\citep{Paturel03} are adopted to derive the rotation curve. The
measured \HI flux is $1.3 \pm 0.4 ~\rm{Jy~km~s^{-1}}$, corresponding
to a mass of $0.7 \pm 0.3 \times10^9 \msun$.
Companion UGC 3797,
which is $5\arcmin$ away to the east, is detected in \HI at the same
angular distance but we do not detect any \HI tidal tails.

\section{Details of Rotation Curves Estimation}
\label{sec:app_RC}

  In this section we investigate two complications that may affect
  rotation curve estimation. First, we consider differences in
  rotation curves arising from different methods of assigning
  velocity. Second, we look at tilted ring modeling, which
  captures warps in the rotation disk, to see if our circular velocities are changed.

  As discussed in \S \ref{sec:fit Vfield}, there are various methods
  to assign velocity to a spectrum, e.g., the peak velocity, the
  intensity-weighted mean (moment-1), or Gaussian/Gauss-Hermite
  velocities. The first two methods have some drawbacks. The
  peak velocity is sensitive to the noise in the data, and is
  discretized to the channel width of the data cube, and the moment-1
  velocity is biased towards the wing if the spectral line is
  asymmetric. Our preferred method is therefore the
  Gaussian/Gauss-Hermite fit. The averaged difference between the
  Gaussian/Gauss-Hermite and the peak velocity (moment-1) velocity field
  is 17 (18) \kms~ for the high SNR galaxies (NGC 2748, NGC 7582, and NGC
  1194) and 37 (33) \kms~ for low SNR galaxies (NGC 2960 and UGC
  3789).

\begin{figure*}[h]
	\centering
	\includegraphics[bb = 80 0 1000 400,clip,scale=0.55]{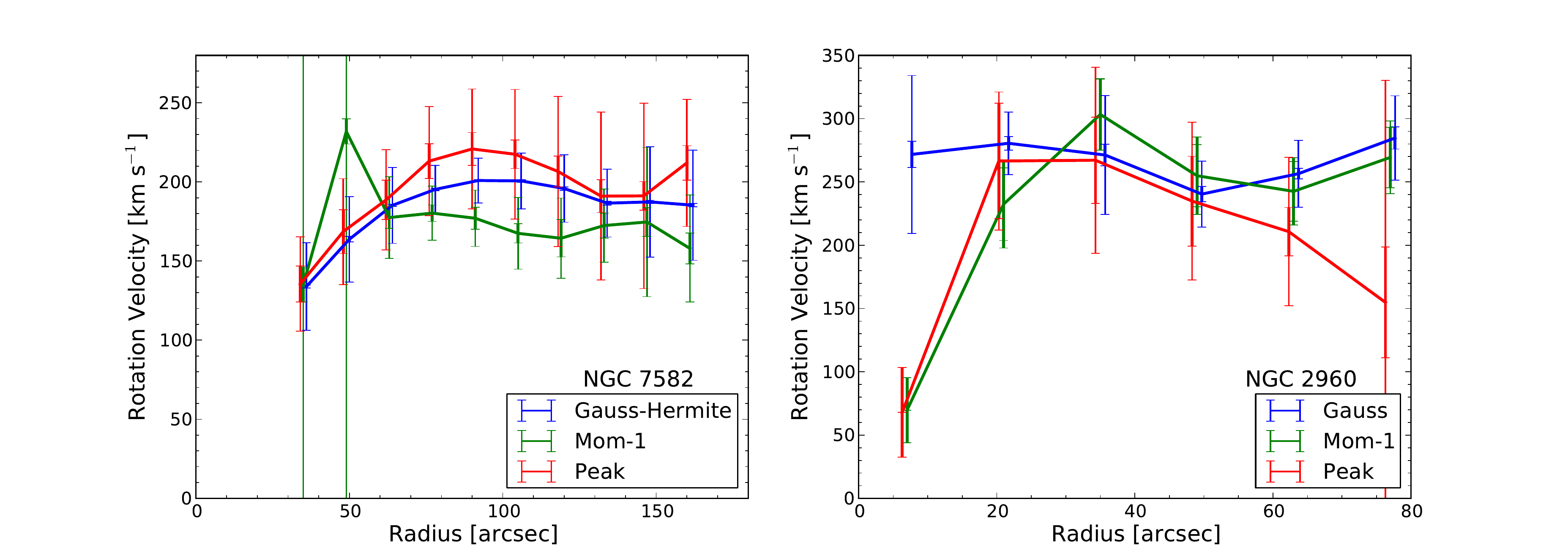}
        \caption{Comparing the rotation curves derived from
            different velocity fields - Gaussian/Gauss-Hermite (blue),
            moment-1 (green), and peak velocity (red). The left is NGC
            7582 with high SNR = 23, and the right is NGC 2960 with
            lower SNR = 6. The thick error bars represent the fitted
            error and the thin error bars are the RMS error. 
            NGC 7582 has $\Vc$, the rotation velocity at large radius, from 
            the three velocity fields consistent with each other. However, NGC 2960 has 
            peak-velocity $\Vc$ significantly differs from the other two methods.
            Therefore, the
            peak velocity may incur large errors in the rotation
            curves for low SNR data. }
\label{fig:compare_RC}
\end{figure*}

We now propagate these different velocity assignments to 
investigate the differences in our inferred rotation curves. We focus on
two representative galaxies, NGC 7582 (high SNR = 23) and NGC 2960
(low SNR = 6). We use the same MCMC
procedure as discussed in \S \ref{sec:fit RC}, assuming a homogeneous
error in the velocity field of 20 \kms~ for NGC 7582 and 40 \kms~ for
NGC 2960, and applying the same masking as in the Gaussian/Gauss-Hermite
velocity field to mask out noisy pixels. The fitted rotation curves
are shown in Fig. \ref{fig:compare_RC}. The average differences
between Gaussian/Gauss-Hermite and the peak velocity (moment-1) rotation
curves are 11 (23) \kms~ for NGC 7582 and 67 (39) \kms~ for NGC
2960. Compared to the RMS errors of the Gaussian/Gauss-Hermite rotation
curves, which are 24 \kms~ for NGC 7582 and 37 \kms~ for NGC 2960, we
found that: First, the peak velocity rotation curve of NGC 7582 is
consistent with the Gaussian/Gauss-Hermite rotation curve within the
error. However, the peak velocity rotation curve deviates from the
Gaussian significantly for the lower SNR galaxy NGC 2960. 

These tests confirm our concern that the peak velocity is highly
sensitive to noise and is not suitable for estimating the rotation
curves for low SNR data. Also, NGC 7582 has a moment-1--derived
rotation curve that is systematically lower than the
Gaussian/Gauss-Hermite one, with a deviation comparable to the RMS
error. This is consistent with our expectation that the moment-1
velocity is biased by the wing of the spectral line. Moment-1 velocities usually
underestimate the rotation velocity for high SNR data, due to beam
smearing and higher intensity at the inner part of the galaxy. This
bias was also demonstrated in the left panel of
Fig. \ref{fig:compare_spec_step}. NGC 2690 has a moment-1 rotation
curve that differs from the Gaussian rotation curve at a level that is
comparable to the RMS errors as well.  In this case we see no systematic
bias, possibly because the SNR is not high enough to manifest an
asymmetric line shape. Considering these effects, we find that the
Gaussian/Gauss-Hermite method is relatively reliable in the face of
asymmetric line shapes and noisy data.

\begin{figure*}[h]
	\includegraphics[bb = 20 570 600 720,clip,scale=1]{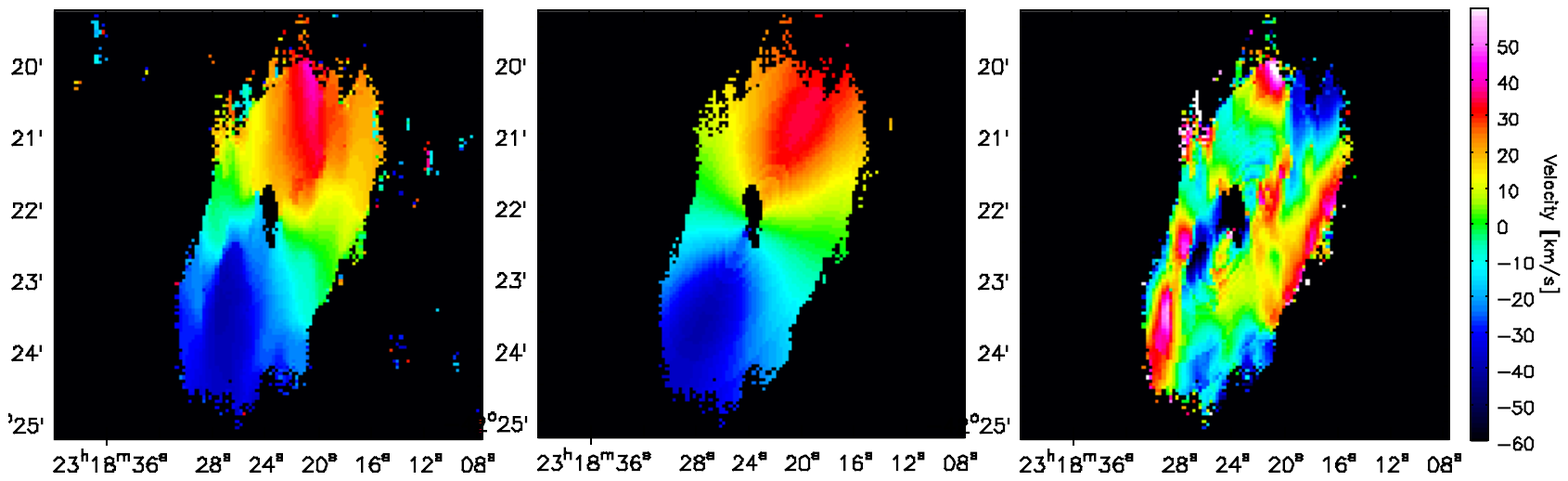}
	\includegraphics[bb = 20 570 600 720,clip,scale=1]{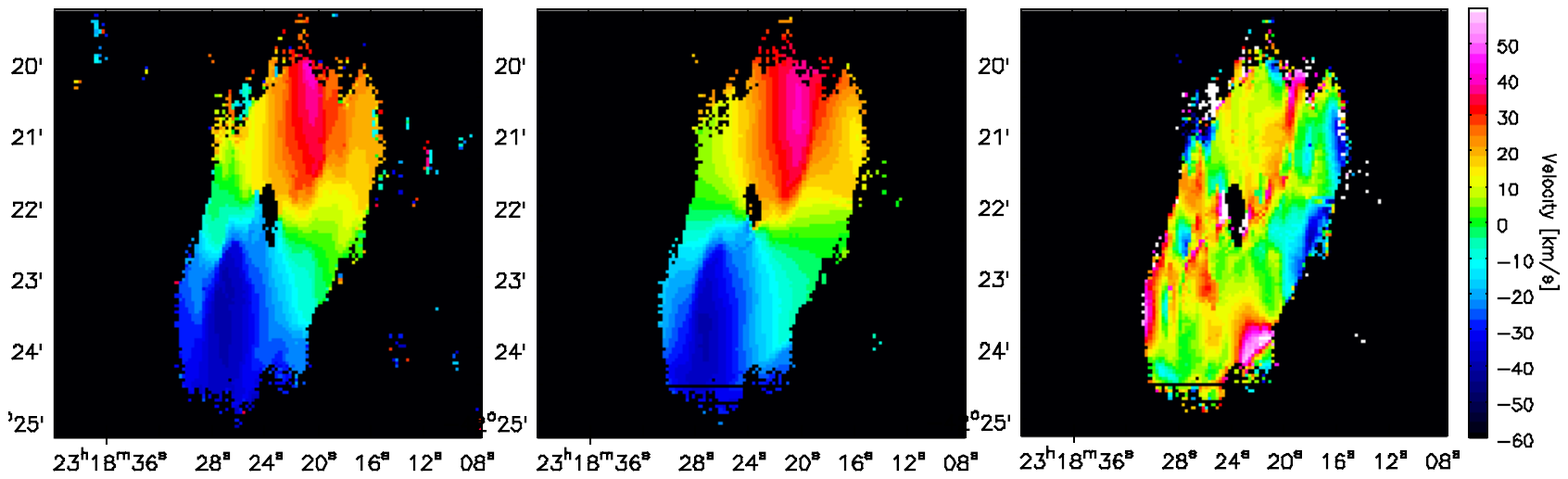}
	\caption{ Comparing coplanar (top) and tilted-ring (bottom) models for the velocity field of NGC 7582. From left to right we show the data (Gauss-Hermite velocity field), the best-fit model, and the residuals. The coplanar model is described in \S \ref{sec:fit RC} and the tilted-ring model in Appendix \ref{sec:app_RC}. The color bar represents the color scheme of the residuals. While the tilted-ring model captures the warping feature in the velocity field, we show in Fig. \ref{fig:TR_RC} that there is very little impact on the inferred rotation curve. }
        \label{fig:TR_model}
\end{figure*}

\begin{figure*}[h]
	\centering
        \hskip 0.8 in
	\includegraphics[scale=0.75]{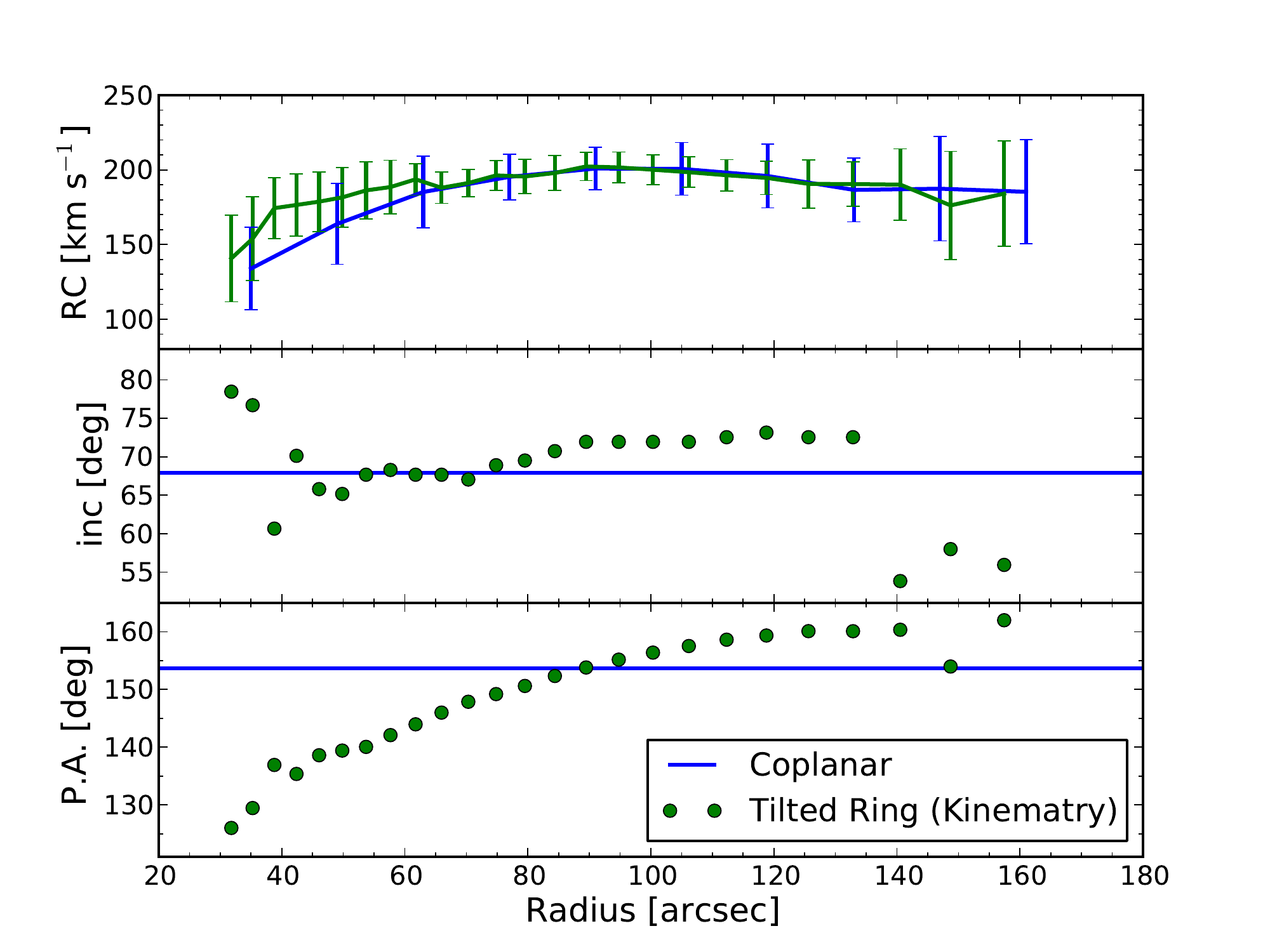}
	\centering
        \caption{The rotation curves (top), inclinations (middle), and position angles (bottom) of the coplanar (blue) and {\it Kinemetry} tilted-ring (green) models for NGC 7582, for details see Appendix \ref{sec:app_RC} and Figure \ref{fig:TR_model} above. The error bars in the rotation curve measurements represent the RMS variation in the residual map. The rotation curves of the two models are consistent with each other within the RMS errors, even when the inclination and P.A. of the tilted ring model fluctuates about the coplanar value. The RMS errors in the tilted-ring models are smaller than in the coplanar model by $40\%$ between 60 and 130$\arcsec$, meaning that some of the variations are accounted for by the higher-order terms and the tilted-rings of {\it Kinemetry}. }
        \label{fig:TR_RC}
\end{figure*}

In \S \ref{sec:fit RC} we adopt a coplanar disk model for rotation
curve fitting. However, this model does not capture the asymmetry in the
velocity field that arises from warping of the disk, as seen in NGC
2748 and NGC 7582. Here we assess whether these asymmetries have an
effect on our $\Vc$ estimates. We take our most asymmetric galaxy NGC
7582 as an example, and use the {\it Kinemetry} method \citep{Kinemetry} to fit a
tilted-ring model. In this model, the position angle and the
inclination of each ring are allowed to vary. Furthermore, second and
third order harmonic terms, e.g., $\sin(2\psi)$ and $\sin(3\psi)$, are
included to capture higher frequency variations along each ring. The
best-fit models are plotted in Fig. \ref{fig:TR_model} with their
residuals, and the fitted rotation curves are plotted in
Fig. \ref{fig:TR_RC}. The tilted-ring model better captures warps
in the disk and reduces the residuals, but the best-fit rotation
velocities stay unchanged within the uncertainties. The difference at
the outer-most bin is less than $5\%$ of the RMS error. This
demonstrates the robustness of the rotation curve fitting against
warps and higher order variations of the velocity field, and shows
that the RMS error is a conservative estimate of the potential
systematics in the rotation curve fitting.

\section{$\Vc$ from Single Dish Measurements}
\label{sec:app_sec}

In our secondary sample we have three galaxies (NGC 3368, NGC 3393,
NGC 3489) with dynamical $\mbh$ but no available rotation curves.
However, $\Vc$ can also be inferred from the line width of integrated
\HI spectra taken with single dish radio observations ($V_{\rm{c},
  \rm{SD}}$), and the single dish $\Vc$ for these three galaxies from
HyperLeda \citep{Paturel03} are listed in Table
\ref{tab:VcReliability_secondary}.

Since single-dish measurements are more readily available for large
samples, they have been used in previous scaling-relation studies
\citep{Pizzella05, Courteau07, Ho07, Beifiori12}. \citet{Roberts78}
and \citet{Ho07} find that $V_{\rm{ c}, \rm{SD}}$ is a robust substitute for
$V_{\rm{c}, \rm{RC}}$.  On the other hand,
without spatial information, single-dish circular velocities may
contain large uncertainties due to the distribution of atomic gas,
irregular rotation-curve shape, or contamination from companion
galaxies.  In general, the values may skew towards lower values, since
single-dish measurements are biased to the inner part of the rotation
curve.

\begin{figure*}[h]
        \hskip 0.8 in
	\includegraphics[bb = 0 0 800 500,clip, scale=0.60]{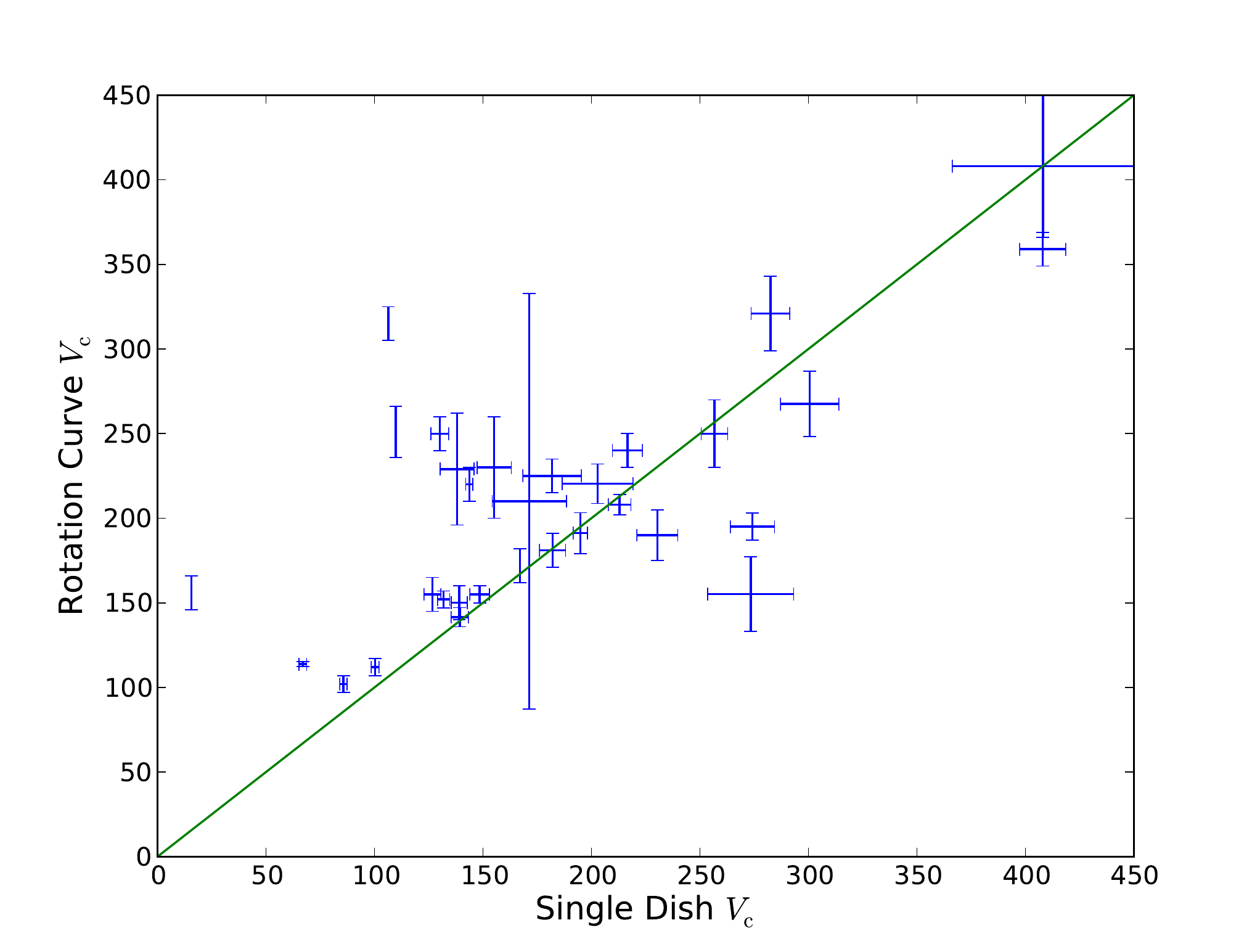}
        \caption{ Single dish circular velocity versus rotation curve
          circular velocity for all galaxies in our sample with both. 
          The solid green line represents $V_{\rm{c}, \rm{SD}} = V_{\rm{c},
            \rm{RC}}$. }
        \label{fig:VcSD-VcRC}
\end{figure*}

To understand how much $V_{\rm{c}, \rm{SD}}$ can deviate from
$V_{\rm{c}, \rm{RC}}$, we compare the two numbers for all galaxies in
our sample that have both measurements (Figure
\ref{fig:VcSD-VcRC}). For most of the galaxies, $V_{\rm{c}, \rm{SD}}$
is consistent with $V_{\rm{c}, \rm{RC}}$. However for a few of them
the two numbers can deviate by up to a factor of two. These are
preferentially S0 galaxies, suggesting that in these cases low gas
fractions are skewing the $V_{\rm{c}, \rm{SD}}$ values low.  On the
other hand, there are only three galaxies in our sample that only have
single-dish \HI measurements. In practice, including or excluding
these three galaxies from our fitting does not change the result. We
decide not to include them in our primary sample.

\section{$\mbh$ Upper Limits}
\label{sec:MBHUL}
\begin{figure*}[h]
	\hskip -0.45 in
	\includegraphics[scale=0.50]{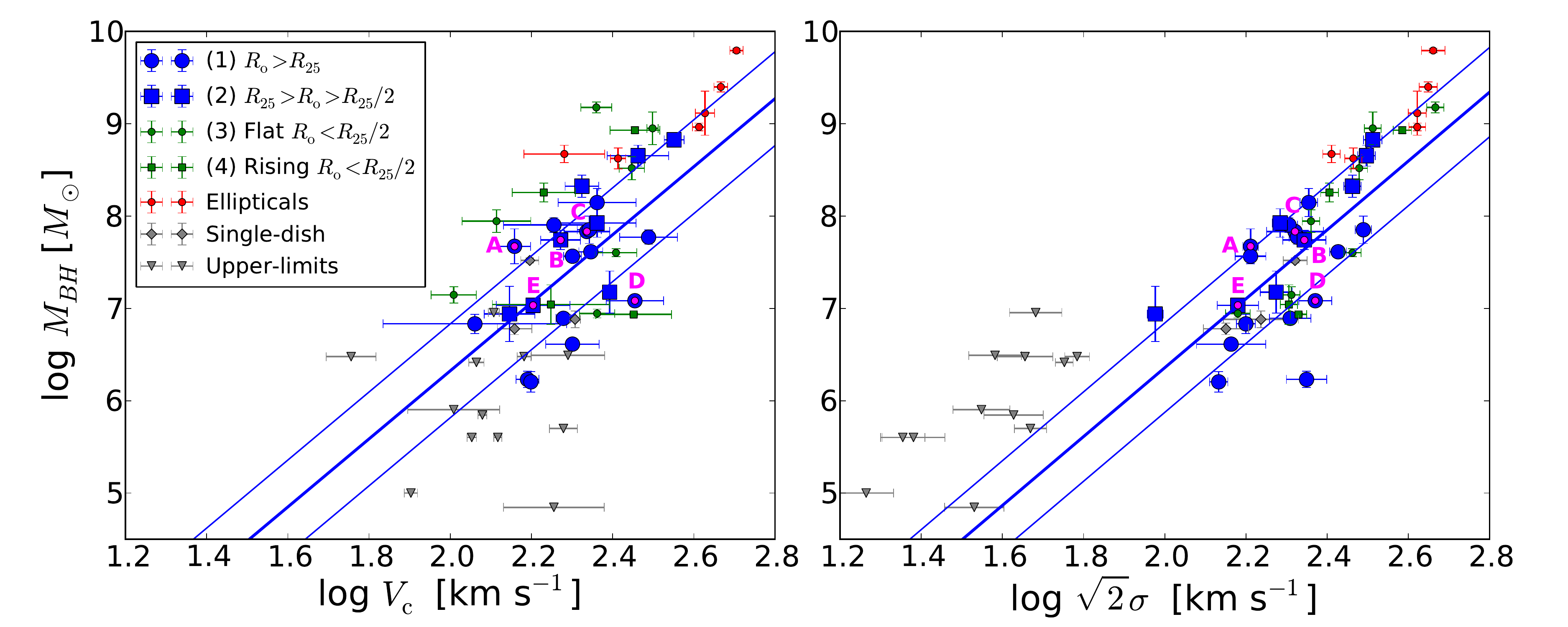}
        \caption{The $\mvc$ (left) and $\msigma$ (right) relations. The data is described in \S \ref{sec:MBH-Vc} and listed in Table \ref{tab:MBHVc}.  Our primary sample with dynamical $\mbh$ and spatially resolved $\Vc$ measurements is plotted in blue or green depending on the rotation curve extent. The blue circles ($R_{\rm{o}} > R_{25}$) and blue squares ($R_{25} > R_{\rm{o}} > R_{25}/2$) have long rotation curves and are used for constraining the two relations. The green circles/squares have short rotation curves ($R_{\rm{o}} < R_{25}$) with flat/rising trends. They are not used to constrain the two relations because of their lower reliability reflected by the larger scatter in the $\mvc$ relation. Also plotted is our secondary sample described in Appendix \ref{sec:app_sec} and listed in Table  \ref{tab:MBHVc_sec}. The grey diamonds and triangles represent the single-dish $\Vc$ measurements and $\mbh$ upper limits respectively. The red circles are the elliptical galaxies with $\Vc$ measured by dynamical modeling and $\mbh$ from \citet{McConnell12} as discussed in \S \ref{sec:fit_results}. Our five observed H~I galaxies are marked by magenta dots labeled as A (NGC 2748), B (NGC 7582), C (NGC 1194), D (NGC 2960), and E (UGC 3789). The fitted $\mvc$ and $\msigma$ scaling relations using the $R_{\rm{o}} > R_{25}/2$ primary sample, plotted in blue, is shown by the thick blue line with the intrinsic scatter plotted by the two thin lines. We use $\sqrt{2}\sigmastar$ for the abscissa of the $\msigma$ relation as a direct comparison to $\Vc$. }
        \label{fig:MBHVc}
\end{figure*}

For completeness, dynamically constrained $\mbh$ upper limits for
bulgeless galaxies, mostly from \citet{Neumayer12}, are listed in the
second section of Table \ref{tab:VcReliability_secondary} and
\ref{tab:MBHVc_sec}. The $\Vc$ values are assigned as described in \S
\ref{sec:Vcsample}, except for the single-dish $\Vc$, which are from
HyperLeda \citep{Paturel03}. These upper limits are also plotted in
Figure \ref{fig:MBHVc} as grey triangles in both of the scaling
relations. Some of them are outliers in the $\mvc$ relation (see
discussion in \S \ref{sec:upper-limits}).

\end{appendix}


\bibliographystyle{apj}

\newpage


\clearpage

\begin{deluxetable}{cclcccccc}
\tablecolumns{9} 
\tabletypesize{\scriptsize}
\tablewidth{0pc}
\tablecaption{VLA Observations \label{tab:obsquality}}
\tablehead{ 
\colhead{Galaxy} & \colhead{Date} & 
\colhead{Flux Cal.} & \colhead{Phase Cal.} & \colhead{$\Delta \theta$} &
\colhead{$T_{\rm{total}}$} & \colhead{$T_{\rm{scan}}$} & \colhead{Antennas} & \colhead{RFI} \\
\colhead{} & \colhead{(UTC)} & 
\colhead{} & \colhead{} & \colhead{(degrees)} & 
\colhead{(minutes)}  &\colhead{(minutes)} & \colhead{} & \colhead{}\\
\colhead{(1)} & \colhead{(2)} & \colhead{(3)} & 
\colhead{(4)} & \colhead{(5)} & \colhead{(6)} & 
\colhead{(7)}  &\colhead{(8)} &\colhead{(9)} 
}
\startdata
  NGC 1194 & 2010 Oct 8 & 3C48 & J0323$+$0534 & 8.3 & 212 & 26.6 & 22 & Yes\\
  NGC 2748 & 2010 Oct 10 & 3C147 & J0841$+$7053 & 6.0 & 217 & 24.6 & 22 & No\\
  NGC 2960 & 2010 Nov 19 & 3C286 & J0943$-$0819 & 11.9 & 210 & 26.3 & 25 & Yes\\
  NGC 7582 & 2010 Dec 4/5 & 3C48 & J2326$-$4027 & 2.5 & 200 & 19.9 & 23 & Yes\\
  UGC 3789 & 2010 Oct 7 & 3C147 & J0614$+$6046 & 8.2 & 198 & 22.0 & 22 & No
\enddata
\tablecomments{Col. (1): Galaxy name. Col. (2): Observation date. Col. (3): The flux and bandpass calibrator. Col. (4): The phase calibrator. Col. (5): The angular separation between the source (galaxy) and the phase calibrator. Col. (6): Total on-source observation time. Col. (7): Average length of each source scan, which is the separation between two phase calibrator scans. Col. (8): Number of antennas used in the observation. Some antennas were not used because the L-band receiver was not yet installed or because the antenna had unstable or noisy data quality. Col. (9): Whether or not radio frequency interference (RFI) was found in the data. The RFI was visually inspected and flagged. After the flagging, there was negligible or minor contamination from RFI in the data cube. The most severe case was NGC 1194, where some faint elongated stripes parallel to the galaxy can be seen. }
\end{deluxetable}

\begin{deluxetable}{ccccccccc}
\tablecolumns{9} 
\tabletypesize{\scriptsize}
\tablewidth{0pc}
\tablecaption{Image Quality \label{tab:image quality}}
\tablehead{ 
\colhead{Galaxy} & \colhead{Distance} & 
\colhead{Channel Width} & \colhead{Weighting} & \colhead{Noise} &
\colhead{Peak} & \colhead{Beam FWHM} & \colhead{Beam P.A.} & \colhead{SNR}\\
\colhead{} & \colhead{(Mpc)}  & 
\colhead{(km s$^{-1}$)} & \colhead{} & \colhead{(mJy beam$^{-1}$)} & 
\colhead{(mJy beam$^{-1}$)}  &\colhead{(arcsec$\times$arcsec)} & \colhead{(degrees)}
& \colhead{}\\
\colhead{(1)} & \colhead{(2)} & \colhead{(3)} & 
\colhead{(4)} & \colhead{(5)} & \colhead{(6)} & 
\colhead{(7)}  &\colhead{(8)} &\colhead{(9)}
}
\startdata
  NGC 2748 & 24.9 & 10 & Robust$=$0.5 & 0.58 & 21.98 & 19$\times$13 & 45 & 38\\
  NGC 7582 & 22.3 & 10 & Robust$=$0.5 & 0.81 & 18.99 & 52$\times$13 & 0 & 23\\
  NGC 1194 & 55.5 & 20 & Natural & 0.41 &5.80 & 23$\times$16 & $-2$ & 14\\
  NGC 2960 & 75.3 & 20 & Natural & 0.37 & 2.29 & 19$\times$18 & $-14$ & 6\\
  UGC 3789 & 48.4 & 20 & Natural & 0.58 & 3.42 & 25$\times$16 & 73 &6
\enddata
\tablecomments{Col. (1): Galaxy name. Col. (2): Distance of the galaxy. For consistency, we adopt distances from \citet{McConnell12}, which is used for $\mbh$ measurements listed in Tab. \ref{tab:MBHVc}. Col. (3): The binned channel (or image plane) width. The channel width of NGC 2748 and NGC 7582 are set to be smaller because of their higher signal-to-noise ratio. Col. (4): The imaging weighting method. Col. (5): The RMS noise level in each image plane measured in line-free channels. Col. (6): The peak intensity in the data cube. Col. (7): The clean beam full-width-half-maximum of the major and minor axes. Col. (8): The clean beam position angle. Col. (9): The signal-to-noise ratio, defined by the maximum intensity per channel divided by the RMS noise. }
\end{deluxetable}

\begin{deluxetable}{cccccc}
\tablecolumns{6} 
\tabletypesize{\scriptsize}
\tablewidth{0pc}
\tablecaption{Fitted disk geometric parameters \label{tab:best fits}}
\tablehead{ 
\colhead{Galaxy} & \colhead{$x_0$} & \colhead{$y_0$} & \colhead{$V_{sys}$} & \colhead{$i$} &
\colhead{P.A.}\\
\colhead{} & \colhead{(R.A.)} & \colhead{(Dec)} & \colhead{(km s$^{-1}$)} & \colhead{(degrees)} & 
\colhead{(degrees)} \\
\colhead{(1)} & \colhead{(2)} & \colhead{(3)} & 
\colhead{(4)} & \colhead{(5)} & \colhead{(6)}
}  
\startdata
  NGC 2748 & $09^{\rm{h}}13^{\rm{m}}43^{\rm{s}}.33$ & $+~76\degree 28\arcmin 35~5\arcsec$ & 1482 & 72.6 &  35.5\\
  NGC 7582 & $23^{\rm{h}}18^{\rm{m}}23^{\rm{s}}.62$ & $-~42\degree 22\arcmin 14~0\arcsec$ & 1588 & 67.9 & 153.7\\
  NGC 1194 & $03^{\rm{h}}03^{\rm{m}}49^{\rm{s}}.14$ & $-~01\degree 06\arcmin 14~8\arcsec$ & 4075 & 69.1 & 142.8\\
  NGC 2960 & $09^{\rm{h}}40^{\rm{m}}36^{\rm{s}}.46$ & $+~03\degree 34\arcmin 36~6\arcsec$ & 4939 & 41.5*&  49.5\\
  UGC 3789 & $07^{\rm{h}}19^{\rm{m}}31^{\rm{s}}.02$ & $+~59\degree 21\arcmin 17~9\arcsec$ & 3229 & 43.2*& 164.7*
\enddata
\tablecomments{The rotating-disk model parameters as described in section ~\ref{sec:fit RC}. Col. (1): Galaxy name. Col. (2,3): Center position. 
For NGC 2748 and NGC 7582 the center position is the fitted H~I kinematical center with fitting error $\pm0.2\arcsec$. For the megamaser galaxies (NGC 1194, NGC 2960, and UGC 3789) the center is fixed at the maser position \citep{Kuo11}. Col. (4): Fitted systemic velocity. The random error from fitting is less than 0.1$\%$, including determining the center for NGC 2748 and NGC 7582, which have no megamaser positions. Col. (5): Fitted inclination angle, except for NGC 2960 and UGC 3789, which are fixed at HyperLeda values \citep{Paturel03}. Col. (6): Fitted position angle, except for UGC 3789, which is fixed at HyperLeda value \citep{Paturel03}. }
\end{deluxetable}

\newpage

\begin{deluxetable}{clccccc}
\tablecolumns{7} 
\tabletypesize{\scriptsize}
\tablewidth{0pc}
\tablecaption{$V_c$ Sources for Primary Sample \label{tab:VcReliability}}
\tablehead{
\colhead{Galaxy} & \colhead{$\Vc$} & \colhead{$\Vc$ Method} & \colhead{$\Vc$ Trend} & \colhead{Inc.} & \colhead{$R_{\rm{o}}/R_{25}$} & \colhead{$\Vc$ Reference} \\
\colhead{} & \colhead{(km s$^{-1}$)} & \colhead{} & \colhead{} & \colhead{(degrees)} & \colhead{} & \colhead{}\\
\colhead{(1)} & \colhead{(2)} & \colhead{(3)} & \colhead{(4)} & \colhead{(5)} & \colhead{(6)} & \colhead{(7)}
}
\startdata
\multicolumn{7}{c}{(Group 1) $R_{\rm{o}} > R_{25}$}\\
\hline
Circinus & 155$\pm$10$\pm$\nodata & HI & flat & 65 & 7.5 & 1\\
Milky Way & 200$\pm$20$\pm$30 & HI & oscillating & \nodata & 1.3 & 2 \\
NGC 0224 & 230$\pm$15$\pm$50 & HI & declining & 77 & 1.7 & 2\\
NGC 1194 & 217$\pm$21$\pm$28 & HI & slightly rising & 69 & 1.8 & \nodata\\
NGC 1300 & 220$\pm$15$\pm$10 & HI & flat & 35 & 1.6 & 3\\
NGC 2273 & 190$\pm$8$\pm$\nodata & HI & flat & 55 & 2.4 & 4\\
NGC 2748 & 144$\pm$12$\pm$13 & HI & flat & 73 & 2.0 & \nodata\\
NGC 2787 & 222$\pm$5$\pm$15 & HI & flat & 42 & 4.3 & 5\\
NGC 2960 & 285$\pm$46$\pm$44 & HI & oscillating & 42 & 2.1 & \nodata\\
NGC 3031 & 180$\pm$10$\pm$50 & HI & slightly declining & 59 & 1.6 & 2\\
NGC 4258 & 200$\pm$8$\pm$10 & HI & slightly oscillating & 67 & 1.8 & 2\\
NGC 4736 & 115$\pm$5$\pm$55 & HI & declining & 47 & 1.9 & 6\\
NGC 4826 & 158$\pm$5$\pm$5 & HI & oscillating & 64 & 1.9 & 6\\
NGC 5128$^a$ & 308$\pm$20$\pm$61& HI& unknown & 47 & 1.2 & 7\\
\hline
\multicolumn{7}{c}{(Group 2) $R_{25} > R_{\rm{o}} > R_{25}/2$}\\
\hline
NGC 3227 & 247$\pm$5$\pm$10 & HI & flat & 56 & 0.8 & 8\\
NGC 3245 & 211$\pm$20$\pm$20 & optical & flat & 62 & 0.6 & 9\\
NGC 4526 & 290$\pm$40$\pm$50 & optical & flat & 90 & 0.6 & 10 \\
NGC 4594 & 356$\pm$20$\pm$20 & HI & flat & 85 & 0.8 & 11 \\
NGC 4596 & 230$\pm$30$\pm$50 & optical & flat & 38 & 0.7 & 12\\
NGC 7457 & 140$\pm$10$\pm$20 & optical & rising & 73 & 0.7 & 9\\
NGC 7582 & 187$\pm$21$\pm$15 & HI & slightly declining & 68 & 0.6 & \nodata\\
UGC 3789 & 160$\pm$33$\pm$10 & HI & rising & 43 & 0.6 & \nodata\\
\hline
\multicolumn{7}{c}{(Group 3) Flat $R_{25}/2 > R_{\rm{o}}$ }\\
\hline
NGC 1023 & 256$\pm$30$\pm$0 & optical & flat & 76 & 0.5 & 13\\
NGC 1332 & 229$\pm$20$\pm$10 & optical & flat & 80 & 0.4 & 13\\
NGC 2549 & 102$\pm$10$\pm$13 & optical & flat& 90 & 0.4 & 13\\
NGC 3115 & 315$\pm$10$\pm$\nodata & optical & flat & 50 & 0.3 & 14\\
NGC 3585 & 280$\pm$20$\pm$\nodata & optical & flat & 90 & 0.2 & 15\\
NGC 4388 & 230$\pm$23$\pm$\nodata & HI & flat & 74 & 0.3 & 16\\
NGC 4564 & 130$\pm$10$\pm$25 & optical & flat & 90 & 0.4 & 17\\
\hline
\multicolumn{7}{c}{(Group 4) Rising $R_{25}/2 > R_{\rm{o}}$}\\
\hline
NGC 1068 & 283$\pm$58$\pm$60 & optical & rising & 46 & 0.5 & 2\\
NGC 3384$^b$ & 177$\pm$16$\pm$56 & optical & rising & 62 & 0.2 & 18\\
NGC 3998$^c$ & 285$\pm$40$\pm$57 & HI & unknown & 70 & 0.5 & 19\\
NGC 4026 & 170$\pm$10$\pm$30 & optical & rising & 90 & 0.5 & 13
\enddata
\tablecomments{
  Circular velocity $\Vc$ for the primary sample with both dynamical $\mbh$ and rotation curve $\Vc$. Galaxies are grouped according to their spatial extension and rotation curve variation as an indication of $\Vc$ reliability. Only Groups (1) and (2) are used to constrain the $\mvc$ relation, while Groups (3) and (4) are excluded, as they may suffer from uncertainties due to short spatial extension. Col. (1): Galaxy name. Col. (2): Circular velocity $\Vc$ of the galaxy with reference in Col. (7). The first error is the observational error. 10$\%$ observational error is assumed if it is not presented in the literature. The second one is the variation of the rotation curve or 20$\%$ if the variation is unknown. $\Vc$ is evaluated at the outermost radius $R_{\rm{o}}$. Col. (3): The observational method of rotation curve measurement. Col. (4): Radial trends in the rotation curve. Col. (5): Inclination angle adopted for the $\Vc$ inclination correction. If the rotation curve is not inclination-corrected in the literature, the optical inclination from HyperLeda \citep{Paturel03} is applied. Col. (6): Ratio between the outermost radius $R_{\rm{o}}$, at which $\Vc$ is evaluated, and the galaxy radius at the B = 25 mag arcsec$^{-2}$ isophote $R_{\rm{25}}$ from RC2 \citep{RC2}. Col. (7): References for $\Vc$: (1) \citet{Jones99}; (2) \citet{Sofue97}; (3) \citet{Lindblad97}; (4) \citet{Noordermeer07}; (5) \citet{Shostak87}; (6) \citet{deBlok08}; (7) \citet{Schiminovich94}; (8) \citet{Mundell95}; (9) \citet{Cherepashchuk10}; (10) \citet{Pellegrini97}; (11) \citet{Bajaja84}; (12) \citet{Kent90}; (13) \citet{Dressler83}; (14) \citet{Bender94}; (15) \citet{Scorza95}; (16) \citet{Guhathakurta88}; (17) \citet{Halliday01}; (18) \citet{Fisher97}; (19) \citet{Knapp85}; (\nodata) is from this paper. Notes on individual galaxies: \\
  $^a$ $\Vc$ is estimated from H~I velocity field as rotation curve is unavailable.\\
  $^b$ The inclination is estimated by the ratio of the minor/major axes ($\cos i = b/a$) from RC3 \citep{RC3}. \\
  $^c$ $\Vc$ was estimated by \citet{Knapp85} to be $285 \pm 40$
  adopting an inclination larger than 70$\degree$, while rotation
  curve is not available. }
\end{deluxetable}

\newpage
\begin{deluxetable}{cccccccc}
\tablecolumns{8} 
\tabletypesize{\scriptsize}
\tablewidth{0pc}
\tablecaption{$\mbh-\Vc$ \label{tab:MBHVc}}
\tablehead{
\colhead{Galaxy} & \colhead{Morphology} & \colhead{Distance} & \colhead{$\mbh$} & \colhead{$\mbh$ Method} & \colhead{$\sigma$} & \colhead{$\Vc$} & \colhead{$\Vc$ Method} \\
\colhead{} & \colhead{} & \colhead{(Mpc)} & \colhead{(\msun)} & \colhead{} & \colhead{(km s$^{-1}$)} & \colhead{(km s$^{-1}$)} & \colhead{} \\
\colhead{(1)} & \colhead{(2)} & \colhead{(3)} & \colhead{(4)} & \colhead{(5)} & \colhead{(6)} & \colhead{(7)} & \colhead{(8)}
}
\startdata
\multicolumn{8}{c}{(Group 1) $R_{\rm{o}} > R_{25}$}\\
\hline
Circinus & S & 4.0 & $1.7^{+0.4}_{-0.3}\times 10^{6}$ & masers & $158^{+18}_{-18}$ & $155\pm 10$ & H~I\\[1.6ex]
Milky Way & S & 0.008 & $4.1^{+0.6}_{-0.6}\times 10^{6}$ & stars & $103^{+20}_{-20}$ & $200\pm 30$ & H~I\\[1.6ex]
NGC 0224 & S & 0.73 & $1.4^{+0.8}_{-0.3}\times 10^{8}$ & stars & $160^{+8}_{-8}$ & $230\pm 50$ & H~I\\[1.6ex]
NGC 1194$^a$ & S0 & 55.5 & $6.8^{+0.4}_{-0.4}\times 10^{7}$ & masers & $148^{+26}_{-22}$ & $217\pm 28$ & H~I\\[1.6ex]
NGC 1300 & S & 20.1 & $7.1^{+3.4}_{-1.8}\times 10^{7}$ & gas & $218^{+10}_{-10}$ & $220\pm 15$ & H~I\\[1.6ex]
NGC 2273$^a$ & S & 26.8 & $7.8^{+0.5}_{-0.5}\times 10^{6}$ & masers & $144^{+18}_{-16}$ & $190\pm 8$ & H~I\\[1.6ex]
NGC 2748$^b$ & S & 24.9 & $4.7^{+2.0}_{-1.9}\times 10^{7}$ & gas & $115^{+5}_{-5}$ & $144\pm 13$ & H~I\\[1.6ex]
NGC 2787 & S0 & 7.5 & $4.1^{+0.4}_{-0.5}\times 10^{7}$ & gas & $189^{+9}_{-9}$ & $222\pm 15$ & H~I\\[1.6ex]
NGC 2960$^a$ & S & 75.3 & $1.21^{+0.07}_{-0.07}\times 10^{7}$ & masers & $166^{+16}_{-15}$ & $285\pm 46$ & H~I\\[1.6ex]
NGC 3031 & S & 4.1 & $8.0^{+2.0}_{-1.1}\times 10^{7}$ & gas & $143^{+7}_{-7}$ & $180\pm 50$ & H~I\\[1.6ex]
NGC 4258 & S & 7.0 & $3.67^{+0.01}_{-0.01}\times 10^{7}$ & masers & $115^{+10}_{-10}$ & $200\pm 10$ & H~I\\[1.6ex]
NGC 4736 & S & 5.0 & $6.8^{+1.6}_{-1.6}\times 10^{6}$ & stars & $112^{+6}_{-6}$ & $115\pm 55$ & H~I \\[1.6ex]
NGC 4826 & S & 7.3 & $1.6^{+0.4}_{-0.4}\times 10^{6}$ & stars & $96^{+5}_{-5}$ & $158\pm 5$ & H~I\\[1.6ex]
NGC 5128 & S0/E & 4.1 & $5.9^{+1.1}_{-1.0}\times 10^{7}$ & stars & $150^{+7}_{-7}$ & $308\pm 61$ & H~I\\[1.6ex]
\hline
\multicolumn{8}{c}{(Group 2) $R_{25} > R_{\rm{o}} > R_{25}/2$}\\
\hline
NGC 3227 & S & 17.0 & $1.5^{+0.5}_{-0.8}\times 10^{7}$ & stars & $133^{+12}_{-12}$ & $247\pm 10$ & H~I\\[1.6ex]
NGC 3245 & S0 & 21.5 & $2.1^{+0.5}_{-0.6}\times 10^{8}$ & gas & $205^{+10}_{-10}$ & $211\pm 20$ & optical\\[1.6ex]
NGC 4526$^c$ & S0 & 16.4 & $4.5^{+1.4}_{-1.0}\times 10^{8}$ & gas & $222^{+11}_{-11}$ & $290\pm 50$ & optical\\[1.6ex]
NGC 4594 & S & 10.0 & $6.7^{+0.5}_{-0.4}\times 10^{8}$ & stars & $230^{+12}_{-12}$ & $356\pm 20$ & H~I\\[1.6ex]
NGC 4596 & S0 & 18.0 & $8.4^{+3.6}_{-2.5}\times 10^{7}$ & gas & $136^{+6}_{-6}$ & $230\pm 50$ & optical\\[1.6ex]
NGC 7457$^b$ & S0 & 12.2 & $8.7^{+5.2}_{-5.2}\times 10^{6}$ & stars & $67^{+3}_{-3}$ & $140\pm 20$ & optical\\[1.6ex]
NGC 7582 & S & 22.3 & $5.5^{+1.6}_{-1.1}\times 10^{7}$ & gas & $156^{+19}_{-19}$ & $187\pm 21$ & H~I\\[1.6ex]
UGC 3789$^a$ & S & 48.4 & $1.08^{+0.06}_{-0.06}\times 10^{7}$ & masers & $107^{+13}_{-12}$ & $160\pm 33$ & H~I\\[1.6ex]
\hline
\multicolumn{8}{c}{(Group 3) Flat $R_{\rm{o}} < R_{25}/2$}\\
\hline
NGC 1023 & S0  & 10.5 & $4.0^{+0.4}_{-0.4}\times 10^{7}$ & stars & $205^{+10}_{-10}$ & $256\pm 30$ & optical\\[1.6ex]
NGC 1332 & S0  & 22.7 & $1.5^{+0.2}_{-0.2}\times 10^{9}$ & stars & $328^{+16}_{-16}$ & $229\pm 20$ & optical\\[1.6ex]
NGC 2549 & S0  & 12.7 & $1.4^{+0.1}_{-0.4}\times 10^{7}$ & stars & $145^{+7}_{-7}$ & $102\pm 13$ & optical\\[1.6ex]
NGC 3115 & S0  & 9.5 & $8.9^{+5.1}_{-2.7}\times 10^{8}$ & stars & $230^{+11}_{-11}$ & $315\pm 10$ & optical\\[1.6ex]
NGC 3585 & S0 & 20.6 & $3.3^{+1.5}_{-0.6}\times 10^{8}$ & stars & $213^{+10}_{-10}$ & $280\pm 20$ & optical\\[1.6ex]
NGC 4388$^d$ & S & 19.8 & $8.8^{+1.0}_{-1.0}\times 10^{6}$ & masers & $107^{+8}_{-7}$ & $230\pm 23$ & H~I\\[1.6ex]
NGC 4564 & S0  & 15.9 & $8.8^{+2.4}_{-2.4}\times 10^{7}$ & stars & $162^{+8}_{-8}$ & $130\pm 25$ & optical\\[1.6ex]
\hline
\multicolumn{8}{c}{(Group 4) Rising $R_{\rm{o}} < R_{25}/2$}\\
\hline
NGC 1068$^b$ & S & 15.4 & $8.6^{+0.3}_{-0.3}\times 10^{6}$ & masers & $151^{+7}_{-7}$ & $283\pm 60$ & optical\\[1.6ex]
NGC 3384 & E  & 11.5 & $1.1^{+0.5}_{-0.5}\times 10^{7}$ & stars & $143^{+7}_{-7}$ & $156\pm 50$ & optical\\[1.6ex]
NGC 3998 & S0  & 14.3 & $8.5^{+0.7}_{-0.7}\times 10^{8}$ & stars & $272^{+14}_{-14}$ & $285\pm 57$ & H~I\\[1.6ex]
NGC 4026 & S0  & 13.4 & $1.8^{+0.6}_{-0.3}\times 10^{8}$ & stars & $180^{+9}_{-9}$ & $170\pm 30$ & optical
\enddata
\tablecomments{Black hole masses, stellar velocity dispersions and circular velocities of our primary sample. These quantities are plotted in Figure \ref{fig:MBHVc} with error bars symmetrized in log space.  Col. (1): Galaxy Name. Col. (2): Morphology. Col. (3): Distance. Col. (4): Black hole mass measured by method in Col. (5):. Col. (6): Stellar velocity dispersion. Col. (2-6) are taken from \citet{McConnell12}, unless otherwise noted. Col. (7): Circular velocity with error taken as the larger one of observational or rotation curve variation error, for more detail see Table \ref{tab:VcReliability}. Col. (8): Method of $\Vc$ observation. Notes on individual galaxies: \\
$^a$ Black hole mass uncertainty of $6\%$, which is dominated by the error of the distance, is adopted as suggested in \citet{Kuo11}. Note that this error is different from what is listed in \citet{McConnell12}. \\
$^b$ These black hole mass measurements are noted in \citet{McConnell12} as complicated and are excluded in their paper. More discussion and the $\mbh$ references can be found in Section \ref{sec:MBH}. \\
$^c$ The black hole mass is from \citet{Davis13} using molecular gas dynamics. \\
$^d$ Same as footnote $a$ but black hole mass uncertainty of $11\%$ is adopted. 
}
\end{deluxetable}

\clearpage

\begin{deluxetable}{ccccccc}
\tablecolumns{7} 
\tabletypesize{\scriptsize}
\tablewidth{0pc}
\tablecaption{Fitted scaling relations \label{tab:M_BH-Vc_results}}
\tablehead{
\colhead{$x$} & \colhead{Sample} & \colhead{Criteria} & \colhead{n} & \colhead{$\alpha$} & \colhead{$\beta$} & \colhead{$\epsilon_{int}$}  \\
\colhead{(1)} & \colhead{(2)} & \colhead{(3)} & \colhead{(4)} & \colhead{(5)} & \colhead{(6)} & \colhead{(7)}
}  
\startdata
$\Vc$ & Group 1, 2 & $R_{\rm{o}}>R_{25}/2$ & 22 & $7.43^{+0.13}_{-0.13}$ & $3.68^{+1.23}_{-1.20}$ & $0.51^{+0.11}_{-0.09}$\\[1.6ex]
$\sqrt{2}\sigma$ & Group 1, 2 & $R_{\rm{o}}>R_{25}/2$  & 22 & $7.48^{+0.11}_{-0.11}$ & $3.73^{+0.87}_{-0.89}$ & $0.48^{+0.10}_{-0.08}$\\[1.6ex]
\hline\\[1.6ex]
$\Vc$ & Group 1 & & 14 & $7.30^{+0.16}_{-0.15}$ & $2.39^{+1.79}_{-1.49}$ & $0.53^{+0.14}_{-0.11}$\\[1.6ex]
$\Vc$ & Group 1, 2, 3 & & 29 & $7.59^{+0.12}_{-0.12}$ & $3.01^{+1.02}_{-0.97}$ & $0.62^{+0.10}_{-0.08}$\\[1.6ex]
$\Vc$ & Group 1, 2, 3, 4 & all Primary Sample & 33 &  $7.61^{+0.12}_{-0.12}$ & $2.86^{+0.99}_{-0.94}$ & $0.65^{+0.1}_{-0.08}$
\enddata
\tablecomments{Best-fit $\mvc$ and $\msigma$ scaling relations with model $\log (\mbh/\msun) = \alpha + \beta \log(x/\rm{200~km s^{-1}})$ of different samples. The first two rows include only galaxies with measured rotation curves extending beyond $R_{\rm{o}} > R_{25}/2$ and are used for our $\mvc$ to $\msigma$ comparison.  Col. (1): The $x$ axis of the relation. Col. (2-3): The sample and the sample criteria used in fitting, see Table \ref{tab:VcReliability}. Col. (4): Number of galaxies in the sample. Col. (5): Best-fit intercept $\alpha$. Col. (6): Slope $\beta$. Col. (7): Intrinsic scatter $\epsilon_{int}$.}
\end{deluxetable}

\begin{deluxetable}{ccccc}
\tablecolumns{5} 
\tabletypesize{\scriptsize}
\tablewidth{0pc}
\tablecaption{HI and Optical Properties \label{tab:HI_properties}}
\tablehead{
\colhead{Galaxy} & \colhead{$F_{\rm{HI}}$} &\colhead{$M_{\rm{HI}}$} & \colhead{M$_B$}& \colhead{$M_{\rm{HI}}/L_{B}$} \\
 & \colhead{(Jy km s$^{-1}$)} &\colhead{($10^9 M_{\odot}$)} & \colhead{(mag)} & \colhead{($M_{\odot}~ L_{\odot B}^{-1}$)}\\
\colhead{(1)} & \colhead{(2)} & \colhead{(3)} & \colhead{(4)} & \colhead{(5)}  
}
\startdata
NGC 2748 & 37.4 $\pm$ 5.4  & 5.47 $\pm$ 1.35 & -20.33 & 0.26 \\
NGC 7582 & 19.6 $\pm$ 2.8 & 2.30 $\pm$ 0.57 & -20.70 & 0.08 \\
NGC 1194 & 7.14 $\pm$ 1.07 & 5.19 $\pm$ 1.30 & -20.36 & 0.24 \\
NGC 2960 & 2.18 $\pm$ 0.28 & 2.92 $\pm$ 0.69 & -20.76 & 0.09 \\
UGC 3789 & 1.26 $\pm$ 0.38 & 0.70 $\pm$ 0.25 & -20.50 & 0.03
\enddata
\tablecomments{The H~I flux, H~I mass, and $M_{\rm{HI}/L_B}$ ratios of our five observed galaxies, described in Section \ref{sec:HI_mass}. Col. (1): Galaxy name. Col. (2): H~I fluxes from our VLA observation. Col. (3): Inferred H~I masses from the flux, adopting distances from \citet{McConnell12} as given in Table \ref{tab:image quality}. A distance error of 10$\%$ is assumed. Col. (4): Absolute B-band magnitude corrected for extinction from HyperLeda \citep{Paturel03}. Col. (5): H~I mass to B-band luminosity ratio, derived from Col. (3) and Col. (4). }
\end{deluxetable}

\begin{deluxetable}{clccccc}
\tablecolumns{7} 
\tabletypesize{\scriptsize}
\tablewidth{0pc}
\tablecaption{$\Vc$ Sources for Secondary Sample \label{tab:VcReliability_secondary}}
\tablehead{
\colhead{Galaxy} & \colhead{$\Vc$} & \colhead{$\Vc$ Method} & \colhead{$\Vc$ Trend} & \colhead{Inc.} & \colhead{$R_{\rm{o}}/R_{25}$} & \colhead{$\Vc$ Reference} \\
\colhead{} & \colhead{(km s$^{-1}$)} & \colhead{} & \colhead{} & \colhead{(degrees)} & \colhead{} & \colhead{}\\
\colhead{(1)} & \colhead{(2)} & \colhead{(3)} & \colhead{(4)} & \colhead{(5)} & \colhead{(6)} & \colhead{(7)}
}
\startdata
\multicolumn{7}{c}{Dynamical $M_{BH}$ with Single dish $\Vc$}\\
\hline
NGC 3368 & 203$\pm$6 & SD& \nodata & 49 & \nodata & 2\\
NGC 3393 & 157$\pm$8 & SD& \nodata & 31 & \nodata & 1\\
NGC 3489 & 144$\pm$14 & SD& \nodata & 65 & \nodata & 2\\
\hline
\multicolumn{7}{c}{Upper-limit $M_{BH}$}\\
\hline
IC 0342 & 190$\pm$15$\pm$\nodata & RC & flat & 25 & 1.6 & 3\\
NGC 0205$^a$ & 26$\pm$3$\pm$5 & V-field & unknown & 90 & 0.2 & 4\\
NGC 0300 & 80$\pm$3 & SD& \nodata & \nodata & \nodata & 2\\
NGC 0428 & 180$\pm$30$\pm$50 & optical & oscillating & 48 & 1.3 & 5\\
NGC 0598 & 135$\pm$13$\pm$35 & RC & rising & 50 & 1.8 & 6\\
NGC 1042 & 57$\pm$8 & SD& \nodata & \nodata & \nodata & 2\\
NGC 2139 & 131$\pm$3 & SD& \nodata & \nodata & \nodata & 2\\
NGC 3423 & 120$\pm$3 & SD& \nodata & \nodata & \nodata & 2\\
NGC 3621 & 152$\pm$2$\pm$6 & RC & flat & 65 & 2.7 & 7\\
NGC 5457 & 195$\pm$10$\pm$40 & RC & declining & 18 & 0.5 & 3\\
NGC 6503 & 116$\pm$2$\pm$5 & RC & flat & 74 & 1.9 & 8\\
NGC 7418 & 128$\pm$4 & SD& \nodata & \nodata & \nodata & 2\\
NGC 7424 & 113$\pm$3 & SD& \nodata & \nodata & \nodata & 2\\
NGC 7793 & 102$\pm$5$\pm$26 & RC & slightly declining & 42 & 1.3 & 7
\enddata
\tablecomments{$\Vc$ reliability for the secondary sample.  Col. (1): Galaxy name. Col. (2): Circular velocity $\Vc$ of the galaxy with reference in col. (7). The first error is the observational error. For rotation curve $\Vc$ the second error is the variation in the rotation curve or 20$\%$ if the variation is unknown. For rotation curves, $\Vc$ is evaluated at the outermost radius $R_{\rm{o}}$. Col. (3): The observational method used to derive $\Vc$. RC stands for spatially resolved rotation curve, and SD for H I single dish observation. Col. (4): Radial trend in the rotation curve. Col. (5): Inclination used for $\Vc$ inclination correction. Col. (6): Ratio between the outermost radius $R_{\rm{o}}$ and the galaxy radius at B = 25 mag arcsec$^{-2}$ isophote $R_{25}$ from RC2 \citep{RC2}. Col. (7): References for $\Vc$: (1) HyperLeda \citep{Paturel03}; (2) \citet{Ho07}; (3) \citet{Sofue97}; (4) \citet{Young97}; (5) \citet{Cherepashchuk10}; (6) \citet{Corbelli00}; (7) \citet{deBlok08}; (8) \citet{Begeman91}. Notes on individual galaxies: \\
$^a$ $\Vc$ is estimated from the H~I velocity field as a rotation curve is unavailable.
}
\end{deluxetable}

\begin{deluxetable}{ccccccccc}
\tablecolumns{9} 
\tabletypesize{\scriptsize}
\tablewidth{0pc}
\tablecaption{$\mbh-\Vc$ Secondary Sample \label{tab:MBHVc_sec}}
\tablehead{
\colhead{Galaxy} & \colhead{Morphology} & \colhead{Distance} & \colhead{$\mbh$} & \colhead{$\mbh$ Method} & \colhead{$\mbh$ Ref.} & \colhead{$\sigma$} & \colhead{$\Vc$} & \colhead{$\Vc$ Method}\\
\colhead{} & \colhead{} & \colhead{(Mpc)} & \colhead{(\msun)} & \colhead{} & \colhead{} & \colhead{(km s$^{-1}$)} & \colhead{(km s$^{-1}$)} & \colhead{}\\
\colhead{(1)} & \colhead{(2)} & \colhead{(3)} & \colhead{(4)} & \colhead{(5)} & \colhead{(6)} & \colhead{(7)} & \colhead{(8)} & \colhead{(9)}
}
\startdata
\multicolumn{9}{c}{Dynamical $\mbh$ with Single dish $\Vc$}\\
\hline
NGC 3368 & S & 10.6 & $7.6^{+1.6}_{-1.5}\times 10^{6}$ & stars & 1 & $122^{+28}_{-24}$ & $203\pm 6$ & SD\\[1.6ex]
NGC 3393 & S & 53.6 & $3.3^{+0.2}_{-0.2}\times 10^{7}$ & masers & 1 & $148^{+10}_{-10}$ & $157\pm 8$ & SD\\[1.6ex]
NGC 3489 & S0 & 12.0 & $6.0^{+0.8}_{-0.9}\times 10^{6}$ & stars & 1 & $100^{+15}_{-11}$ & $144\pm 14$ & SD\\[1.6ex]
\hline
\multicolumn{9}{c}{Upper-limit $\mbh$}\\
\hline
IC 0342 & S & 1.8 & $ < 5\times 10^{5}$ & stars & 2 & $33\pm3$ & $190\pm 15$ & RC\\[0.7ex]
NGC 0205 & E (dwarf) & 0.74 & $ < 2\times 10^{4}$ & stars & 3 & $39\pm6$ & $26\pm 5$ &V-field\\[0.7ex]
NGC 0300 & S & 2.2 & $ < 1\times 10^{5}$ & stars & 4 & $13\pm2$ & $80\pm 3$ & SD\\[0.7ex]
NGC 0428 & S & 16.1 & $ < 7\times 10^{4}$ & stars & 4 & $24\pm4$ & $180\pm 50$ & optical \\[0.7ex]
NGC 0598 & S & 0.8 & $ < 2\times 10^{3}$ & stars & 5 & $19.8\pm0.8$ & $135\pm 35$ & RC\\[0.7ex]
NGC 1042 & S & 18.2 & $ < 3\times 10^{6}$ & stars & 4 & $32\pm5$ & $57\pm 8$ & SD\\[0.7ex]
NGC 2139 & S & 23.6 & $ < 4\times 10^{5}$ & stars & 4 & $17\pm3$ & $131\pm 3$ & SD \\[0.7ex]
NGC 3423 & S & 14.6 & $ < 7\times 10^{5}$ & stars & 4 & $30\pm5$ & $120\pm 3$ & SD \\[0.7ex]
NGC 3621 & S & 6.6 & $ < 3\times 10^{6}$ & stars & 6 & $43\pm3$ & $152\pm 6$ & RC\\[0.7ex]
NGC 5457 & S & 7.0 & $ < 3\times 10^{6}$ & stars & 7 & $27\pm4$ & $195\pm 40$ & RC\\[0.7ex]
NGC 6503 & S & 5.27 & $ < 3\times 10^{6}$ & stars & 7 & $40\pm2$ & $116\pm 5$ & RC\\[0.7ex]
NGC 7418 & S & 18.4 & $ < 9\times 10^{6}$ & stars & 4 & $34\pm5$ & $128 \pm 4$ & SD \\[0.7ex]
NGC 7424 & S & 10.9 & $ < 4\times 10^{5}$ & stars & 4 & $16\pm2$ & $113\pm 3$ & SD \\[0.7ex]
NGC 7793 & S & 3.3 & $ < 8\times 10^{5}$ & stars & 4 & $25\pm4$ & $102\pm 26$ & RC
\enddata
\tablecomments{Black hole masses, stellar velocity dispersions, and circular velocities of our secondary sample for either H~I single dish $\Vc$ or upper-limits for $\mbh$. These quantities are plotted in Figure \ref{fig:MBHVc} in grey with error bars symmetrized in log space.  Col. (1): Galaxy Name. Col. (2): Morphology. Col. (3): Distance. Col. (4): Black hole mass measured by method col. (5): from reference col. (6). Col. (7): Stellar velocity dispersion. For the first section (single-dish $\Vc$) Col. (2-7) are taken from \citet{McConnell12} and original references can be found therein. For the second section ($\mbh$ upper-limits) Col. (2-3) are from \citet{McConnell12} and Col. (4, 5, 7) are from the $\mbh$ references listed in Col. (6). Col. (8): Circular velocity with error taken as the larger one of observational or RC variation error (see Table \ref{tab:VcReliability_secondary}). Col. (9): The observational method of $\Vc$. SD stands for spatially unresolved single dish $\Vc$, RC for H~I rotation curve $\Vc$, H~I V-field for spatially resolved data but unavailable rotation curves, and optical for optical rotation curves. References: (1) \citet{McConnell12}; (2) \citet{Boeker99}; (3) \citet{Valluri05}; (4) \citet{Neumayer12}; (5) \citet{Gebhardt01}; (6) \citet{Barth09}; (7) \citet{Kormendy10}
}
\end{deluxetable}

\end{document}